\documentclass{article}
\usepackage[english]{babel}
\usepackage[margin=0.5in]{geometry}
\usepackage{multicol}
\usepackage{background}

\usepackage{amsmath,amssymb}	% to get all AMS symbols
\usepackage{graphicx}			% to insert figures
\usepackage{here}                             % for figure placement

\SetBgScale{4}
\SetBgColor{gray}
\SetBgAngle{90}
\SetBgContents{Draft}
\SetBgPosition{-1.5,-10}
\begin{document}
 
{\centering
 
{\bfseries\Large Diversity of knot solitons in liquid crystals manifested by linking of preimages in torons and hopfions \bigskip}
 
P. J. Ackerman\textsuperscript{1,2} and I. I. Smalyukh\textsuperscript{1,2,3,4} \\
   {\itshape
\textsuperscript{1}Department of Physics, University of Colorado, Boulder, Colorado 80309, USA \\
\textsuperscript{2}Department of Electrical, Computer and Energy Engineering, University of Colorado, Boulder, Colorado 80309, USA \\
\textsuperscript{3} University of Colorado, Liquid Crystal Materials Research Center and Materials Science and Engineering Program, Boulder, CO 80309, USA \\
\textsuperscript{4} Renewable and Sustainable Energy Institute, National Renewable Energy Laboratory and University of Colorado, Boulder, Colorado 80309, USA \\
\normalfont (Dated: May 9, 2016)
 
   }
}
 
\begin{abstract}
Topological solitons are knots in continuous physical fields classified by non-zero Hopf index values. Despite arising in theories that span many branches of physics, from elementary particles to condensed matter and cosmology, they remain experimentally elusive and poorly understood. We introduce a method of experimental and numerical analysis of such localized structures in liquid crystals that, similar to the mathematical Hopf maps, relates all points of the medium's order parameter space to their closed-loop preimages within the three-dimensional solitons. We uncover a surprisingly large diversity of naturally occurring and laser-generated topologically nontrivial solitons with differently knotted nematic fields, which previously have not been realized in theories and experiments alike. We discuss the implications of the liquid crystal's non-polar nature on the knot soliton topology and how the medium's chirality, confinement and elastic anisotropy help to overcome the constrains of the Hobart-Derrick theorem, yielding static three-dimensional solitons without or with additional defects. Our findings will establish chiral nematics as a model system for experimental exploration of topological solitons and may impinge on understanding of such nonsingular field configurations in other branches of physics, as well as may lead to technological applications
\bigskip
 
%\noindent DOI: 10.
 
\end{abstract}

\begin{multicols}{2}

\section{Introduction}

When proposing one of the early models of atoms, Kelvin and Tait considered knotted vortices as candidates and attempted to explain the diversity of chemical elements in the periodic table as that arising from a large variety of knots, giving origins to the modern field of mathematical knot theory [1-3]. Even before them, Gauss envisaged that spatially localized knots in physical field lines, such as magnetic or electric field lines, could behave like particles [1], provided that the crossing of the field lines is prevented, say, due to energetic reasons. Hopf later rigorously demonstrated that, indeed, inter-knotted closed loops could be smoothly embedded in a uniform far-field background, introducing the celebrated mathematical Hopf fibration [1,3,4]. Finkelstein applied these mathematical concepts and Hopf mappings to three-dimensional (3D) physical fields [5], so that the 3D topological solitons based on them subsequently started attracting interest (mostly theoretical) in many branches of physics [6-22]. However, the experimental realizations and demonstrations of topological solitons with knotted field lines typically deal only with transient phenomena and out-of-equilibrium systems or are accompanied by observation of additional defects [3,6-22] and their detailed explorations are often hindered by the need of 3D spatial imaging of the physical fields. Moreover, according to the Hobart-Derrick theorem [7,8], physical systems cannot host the static 3D solitons in continuous fields described within the simplest field theories because of energetic reasons, except for within the nonlinear theories with higher-order derivatives, such as the Skyrme-Faddeev model [6,9,10]. Thus, not surprisingly, the challenge of reliable experimental realization and robust control of topological solitons persisted for decades.

Since the 3D topological solitons smoothly embed into the uniform far-field background, their solitonic field configurations in the 3D space can be effectively ``compactified'' to a 3-sphere $S^3$ and the field topology is then characterized by the $S^3 \rightarrow S^2$ maps, bringing about the beautiful analogy with the famous mathematical Hopf and Seifert fibrations [1,3,4]. Within the homotopy theory classification of topological defects, the 3D solitons in vector fields, with a two-dimensional sphere $S^2$ as the ground-state manifold, belong to the third homotopy group $\pi_3(S^2)=\mathbb{Z}$ and are classic examples of nonsingular topological defects. In the nonpolar systems with line fields, such as the director field in nematic liquid crystals (LCs) [23] that describes the spatial changes of the local average orientation of rod-like molecules, the ground-state manifold is a projective plane $\mathbb{R}P^2$ (or, equivalently, the sphere with antipodal points identified, $S^2/Z_2$), so that the corresponding 3D topological solitons are labeled as $\pi_3(\mathbb{R}P^2)$ or $\pi_3(S^2/Z_2)=\mathbb{Z}$ [14,22,23]. An intriguing and important feature of these topological solitons is that they have closed-loop preimages (regions within the 3D sample that have the same orientation of the physical field) corresponding to all points of the order parameter space (e.g. $S^2$ for the vector fields) and that each two distinct preimages are linked with each other an integer number of times. This linking number is the so-called Hopf index topological invariant, $Q$, characterizing the topology of the 3D topological solitons. 

The scientific and technological potential of 3D topological solitons can be appreciated by considering the recent advances in the studies of their two-dimensional counterparts called ``skyrmions'' or ``baby skyrmions'' [24-30], which belong to the second homotopy group $\pi_2(S^2)=\mathbb{Z}$. Being a subject of purely theoretical studies a decade ago or so [24], they have been recently successfully realized in experimental systems, both in the form of isolated individual solitons and as novel phases with arrays of such solitons in the ground-state [25-30]. These skyrmions attract a great deal of fundamental interest and define foundations for skyrmionics and other emerging technologies [25-30], albeit their experimental study is still largely restricted to chiral condensed matter systems such as non-centrosymmetric ferromagnets and chiral nematic LCs [29,30]. One condensed matter system that has been considered for experimental realization of the 3D topological soliton structures with non-zero Hopf index values is a cholesteric LC. Short-pitch cholesteric LCs are topologically similar to the still elusive biaxial nematic LCs [31,32] and are characterized by three mutually orthogonal nonpolar director fields [23]. Some of the transient localized director structures from experimental observations [33] could be potentially interpreted as being nonsingular in one of these three director fields while also having nonzero Hopf invariants [31], albeit unambiguous demonstration of this topological nature of such structures remained impossible due to the lack and limitations of 3D director field imaging capabilities. Furthermore, such localized field configurations were found only as transient structures [33] and could be potentially nonsingular only in one out of the three director fields of this rather complex condensed matter system [31-33], thus being only loosely related to the hopfions envisaged to exist in many other branches of physics [9,10], albeit realization of various localized structures in biaxial condensed matter systems is also of great fundamental interest [23,34]. With the advent of 3D director imaging capabilities [13,14,35], the reconstruction of complex field configurations became possible, but only a small variety of spatially localized 3D solitonic structures have been found so far in LCs and chiral ferromagnets [13, 14, 22, 36-40]. Furthermore, the understanding of topology, structural diversity and physical underpinnings behind the stability of such 3D solitons is still limited and calls for new experimental and theoretical approaches in their exploration. Most importantly, there is a need for the direct experimental characterization of linking of preimages and Hopf indices corresponding to different 3D solitonic structures.

In this work, we introduce a method of direct experimental and numerical characterization of preimages of the 3D solitons. We then realize and study a series of stationary 3D solitons in a confined chiral nematic LC [23] with the nonpolar director field $\bf n(r)$ describing spatial changes of the local average molecular alignment direction of the constituent rod-shaped molecules. By using a combination of a direct 3D nonlinear optical imaging and numerical modeling through minimization of the free energy that both yield 3D $\bf n(r)$ spatial patterns, we construct the soliton preimages corresponding to all distinct points of the order parameter space. From a large variety of experimentally realized solitonic structures, we focus on solitons with inter-linked preimages in the form of closed loops. These solitons are characterized by preimage linking numbers and the corresponding nonzero topological Hopf index invariants $Q$, different from the 3D localized field configurations of elementary torons with $Q=0$ that we studied previously [13]. Numerical modeling provides insights into the role of the medium's chirality, confinement, and elastic constant anisotropy in enabling the stability of these 3D solitons. We discuss how a combination of these factors helps to overcome the constraints of the Hobart-Derrick theorem [7,8] and how our findings may provide insights into the prospects of obtaining stable topological solitons in other branches of physics, both within condensed matter and well beyond it. Furthermore, the experimental platform we have developed may lead to technological applications building on the particle-like nature of topological solitons as well as to the realization of topological solitonic condensed matter phases.

\section{Theoretical Foundations}

\subsection{3D soliton topology}

The ground-state manifold of a physical system with a characteristic vector quantity as an order parameter, such as the magnetization in a ferromagnet, is a two-dimensional unit sphere $S^2$. An important property of the $\pi_3(S^2)=\mathbb{Z}$ topological solitons with nonzero Hopf invariant, such as hopfions, is that the preimages of all points on $S^2$ are closed loops (Fig. 1a), which are topologically equivalent to one-spheres $S^1$. A preimage of the north-pole point of $S^2$ closes into an $S^1-$like loop through infinity and is typically associated with a uniform far-field background embedding the soliton. The Hopf index invariant can be determined as the linking number of any two preimages of two distinct points on $S^2$ and also describes how many times $S^2$ is swept in the process of mapping the vector field from the 3D space of the soliton (and the corresponding $S^3$) to $S^2$. The topological Hopf index of a soliton is commonly found as the linking number $Q=\sum C/2$ of preimages of any two distinct points on the target $S^2$ (Fig. 1) in the case of vectorized $\bf n(r)$, where the sign of crossings $C=\pm1$ depends on the directions of consistently determined circulations of the soliton's preimages. In the case of nematic LCs with nonpolar symmetry of the director field $\bf n(r)\equiv -\bf n(r)$, the ground-state manifold $S^2/Z_2$ is a sphere with the diametrically opposite points identified. The implication of the LC's nonpolar symmetry is that the preimages with all single orientations of $\bf n(r)$ cannot be distinguished from those with $-\bf n(r)$ (Fig. 1b). Since the director field can be vectorized and since there is a theorem stating that all nonsingular $\pi_3(S^2/Z_2)$ structures in the director field are also nonsingular in the two co-linear mutually opposite vector fields decorating it [31], it is often possible to determine the Hopf index invariant from the linking of the two closed-loops of a single preimage corresponding to a single point on $S^2/Z_2$ (which would correspond to the preimages with two single orientations $\bf n$ and $-\bf n$ in the anti-parallel vector fields decorating the director field, as shown in Fig. 1b). To explore the implications of the nonpolar nature of the director field $\bf n(r)$ on the 3D soliton topology and to compare the solitons in director and vector fields, we use a color scheme for $S^2/Z_2$ that is consistent with the identification of its diametrically opposite points (Fig. 1c). We illustrate the spatial pattern of the director with the help of appropriately colored double cones (Fig. 1c), as opposed to the single cones that we use for visualizing vectors and the vectorized $\bf n(r)$.

The 3D space within a topological soliton is smoothly filled with closed loop preimages that reside on nested tori (Fig. 1d,e). For the elementary solitons in vector fields or in vectorized $\bf n(r)$ (Fig. 1d), the torus circular axis and the far-field background correspond to the south pole and the north pole of $S^2$, respectively. In the case of a director field $\bf n(r)\equiv -\bf n(r)$ with a nonpolar nature, both the far-field and the circular axis of the soliton have the same up-down orientation parallel to the vertical z-axis (Fig. 1e). For the simplest solitons, the topological Hopf invariant $Q$ can be found as a linking number of preimages of any two distinct points on $S^2$ (Fig. 1d) in the case of vectorized $\bf n(r)$ or, equivalently, as the linking 

\begin{figure}[H]
\includegraphics[width=0.49\textwidth]{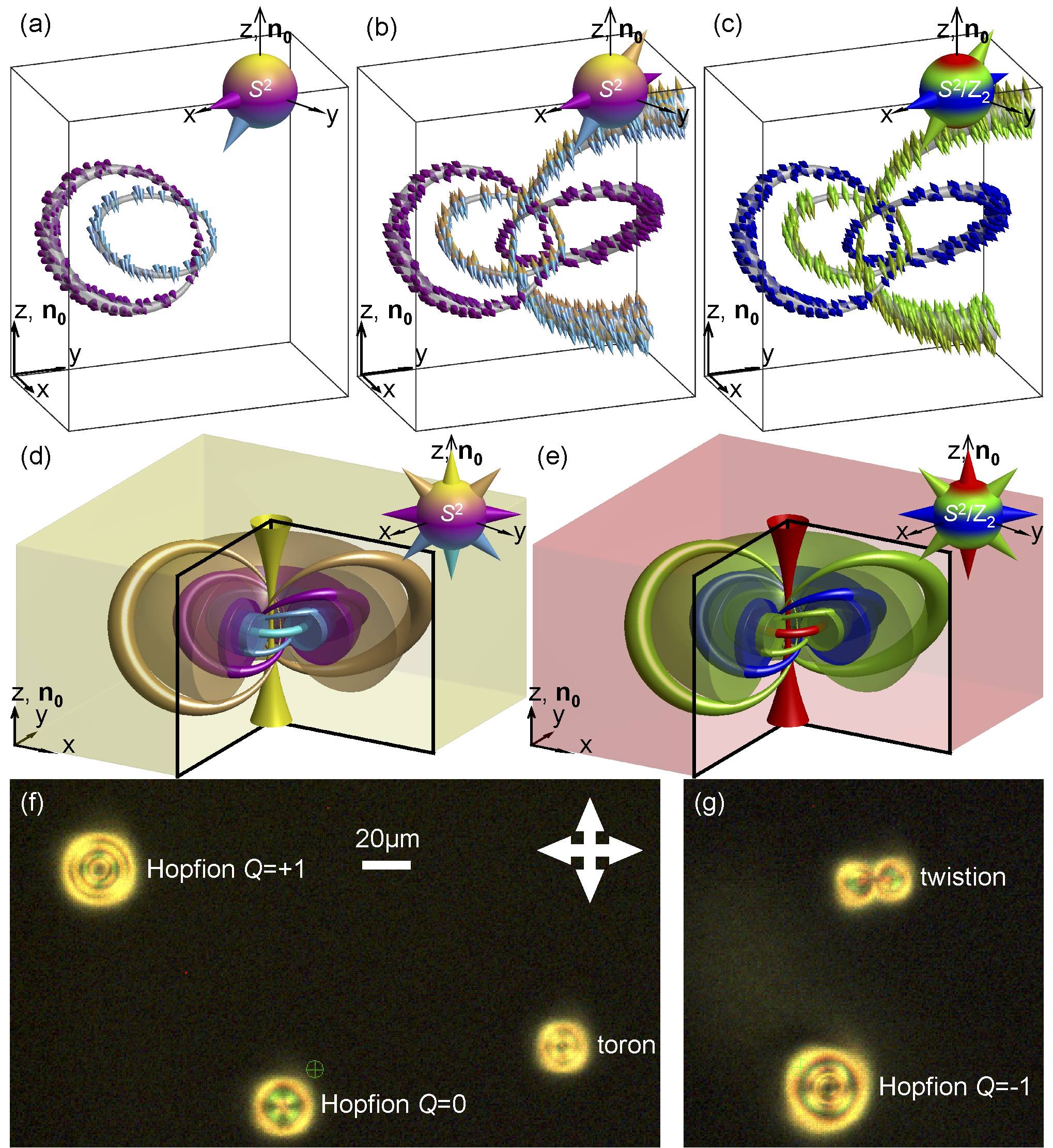}
\caption{3D topological solitons in chiral nematic LCs. (a-c) Examples of closed-loop preimages of 3D topological solitons corresponding to (a) the two distinct points on $S^2$ (shown using cones) for vectorized $\bf n(r)$ and (b,c) two distinct points on $S^2/Z_2$ and nonpolar $\bf n(r)$, with the ground-state manifold $S^2/Z_2$ depicted as a sphere with diametrically opposite  points identified through coloring and making the $\bf n$ and $-\bf n$ orientations non-distinguishable. Note that the diametrically opposite points in (b) appear simultaneously, so that the two points on $S^2/Z_2$ are four points on $S^2$ for the vectorized director field. (d,e) Illustrations of Hopf maps of closed-loop preimages of 3D topological solitons embedded in a uniform far-field $\bf n_0$ onto the (d) $S^2$ ground-state manifold for vectorized $\bf n(r)$ and (e) $S^2/Z_2$ order parameter space of the LC for nonpolar $\bf n(r)$. The schematics show linking of the hopfion's circle-like preimages that reside on nested tori in the sample's 3D space and correspond to color-coded points (cones) on $S^2$ or $S^2/Z_2$. (f,g) Polarizing optical micrographs of the studied localized field configurations co-existing within the same LC sample. The images were obtained between crossed polarizers (white double arrows). The green circles with crosses visible in the micrographs show locations of ``invisible'' infrared laser traps used to manipulate the naturally occurring localized field configurations. The micrographs were obtained for a 5CB-based partially polymerizable nematic LC mixture with chiral additive cholesteryl pelargonate, $d=10\mu$m and $d/p\sim1$.}
\end{figure}

\noindent number of two loops that comprise a preimage of a single point on $S^2/Z_2$ in the case of analyzing the nonpolar $\bf n(r)\equiv -\bf n(r)$ field (Fig. 1e). However, we will also show that there are more complex solitons that require a detailed analysis of preimages with complex linking.

\subsection{Free energy and stability}

For chiral nematic LCs with helicoidal pitch $p$, the elastic free energy cost for producing spatial deformations of $\bf n(r)$ reads [13,23]

\begin{multline}
F = \int dr\bigg \{\frac{K_{11}}{2}(\nabla \cdot \mathbf n)^2+\frac{K_{22}}{2}[\mathbf n \cdot (\nabla \times \mathbf n)]+\frac{K_{33}}{2}[\mathbf n \times (\nabla \times \mathbf n)]^2\\  
+K_{22}q_0\mathbf n \cdot (\nabla \times \mathbf n) - 
K_{24}[\nabla \cdot [\mathbf n (\nabla \cdot \mathbf n) + \mathbf n \times (\nabla \times \mathbf n)]]\bigg \}
\end{multline}

\noindent where $q_0=2\pi/p$ characterizes the LC chirality and the Frank elastic constants $K_{11},~K_{22},~K_{33}$ and $K_{24}$ describe the energetic costs of the splay, twist, bend, and saddle-splay elastic deformations, respectively [23]. Within the $K_{11}=K_{22}=K_{33}=K$ one-constant approximation and while neglecting the energetic cost due to the saddle-splay term and constant terms, Eq. (1) reduces to the form

\begin{equation}
F = \int dr\bigg \{J(\nabla\mathbf n)^2+D\mathbf n \cdot(\nabla \times \mathbf n)\bigg \}
\end{equation}

\noindent where $J=K/2$ and $D=Kq_0$. The free energy within the one-constant approximation is analogous to one of the most common forms of the Hamiltonian describing solid-state chiral ferromagnets, where coefficients $J$ and $D$ in this case describe the strengths of exchange energy and the Dzyaloshinskii-Moriya coupling, respectively [23,24,41]. In the case of $D=0$ (for the non-chiral systems), Eq. (2) further reduces to the Hamiltonian in a form that was considered by Hobart and Derrick [7,8] to demonstrate instability of the localized nonsingular 3D field configurations within such a model. For the solitons to be stable in chiral nematic LCs, they need to emerge as local or global minima of the free energy given by Eqs. (1) and (2) [23]. In addition to chirality, which tends to stabilize the twisted solitonic structures and may help to overcome the constrains of the Hobart-Derrick theorem [7,8], the anisotropy of elastic constants in Eq. (1), confinement and the surface anchoring at confining surfaces may serve a similar role. Chiral nematic LCs are dielectric and diamagnetic materials that respond to external fields to minimize the corresponding terms of the free energy. For example, in the case of the applied electric fields the Eqs. (1) and (2) need to be supplemented by adding the corresponding electric field coupling term of the free energy

\begin{equation}
F_{electric} = -(\epsilon_0 \Delta \epsilon/2)\int dr (\mathbf E \cdot \mathbf n)^2
\end{equation}

\noindent where $\bf E$ is the applied electric field, $\epsilon_0$ is the permittivity of free space, and the dielectric anisotropy $\Delta \epsilon$ can be both positive and negative, depending on the LC material system. This coupling term can potentially also promote stabilization of solitonic structures.

The analytical treatment of Eqs. (1) and (2) from the standpoint of view of stability of complex 3D solitonic structures is prohibitively difficult, especially when supplemented by field coupling terms such as the one given by Eq. (3), and we therefore resort to the numerical minimization of free energy to arrive at a host of field configurations corresponding to local and global energy minima. We note that the free energy of chiral nematic LCs could be also described within the Landau-de Gennes tensorial approach [23,42], which is a method of choice in the cases of structures with singular line defects [42, 43], but the Frank-Oseen free energy description is more suitable for modeling nonsingular field configurations such as the hopfions as well as torons with the additional point singularities [13]. This approach not only allows us to do the full treatment of anisotropic elastic properties of LCs, explore the implications of the one-constant-approximation and the relevance to systems such as chiral ferromagnets, but also to operate with large sizes of grids and samples in the numerical modeling, which is critically important for elucidating the numerical results we present. Furthermore, this Frank-Oseen approach and modeling based on Eqs. (1) and (2) also allow us to better connect our work to the efforts of observing topological solitons in other branches of physics.

\section{Materials and Methods}

\subsection{Materials and sample preparation}

To assure a broad impact of our work and its accessibility to other researchers, we use commercially available LC materials pentylcyanobiphenyl (5CB, Frinton Laboratories, Inc.) and ZLI-2806 (EM Chemicals). To obtain chiral nematic LCs, these nematic hosts are doped with small amounts of chiral additives, either the left-handed dopant cholesteryl pelargonate (Sigma-Aldrich) or the right-handed chiral dopant CB-15 (EM Chemicals). The chiral additive is added to the nematic host at a weight fraction calculated as $c=(\xi\cdot p)^{-1}$, allowing us to define the helicoidal pitch $p$ of the ensuing chiral LC, where $\xi$ is the helical twisting power of the additive in the nematic host [30,36]. The obtained equilibrium pitch value is then probed using the Gradjean-Cano method [13,22]. In addition, a polymerizable nematic mixture of 5CB (69\%) with 12\% of RM-82 and 18\% of RM-257 reactive diacrylate nematics and 1\% Irgacure 184 photoinitiator (all from CIBA Specialty Chemicals) [38] is also doped with the same two chiral agents to obtain partially polymerizable chiral nematic LCs. The initial powder mixture is first dissolved in dichloromethane to homogenize, kept at an elevated temperature of 85 $^\circ$C for one day to remove the solvent, and cooled down to obtain the final chiral nematic mixture. We have optimized the polymerization process by using relatively low light exposures for cross-linking of the cholesteric films, so that the 3D structures of torons and hopfions can be ``frozen'' by polymerization in a solid film without altering their director configurations [38]. The polymerization is achieved using relatively weak ultraviolet exposure by means of a home-built setup with a 20 W mercury bulb (obtained from Cinch) [38]. The unpolymerized 5CB within this partially cross-linked system is partially removed by addition of isopropanol and subsequently replaced with immersion oil. This allows us to reduce the medium's effective birefringence by about an order of magnitude (estimated to be $\approx$0.02) without disrupting the director structure of this partially polymerized sample [38], which is key for the mitigation of depolarization and defocusing effects during the polarized nonlinear optical imaging described below [13,38].

Chiral LC samples with uniform far-field director $\bf n_0$ orientation were prepared by sandwiching the LC mixtures between glass plates with well-defined perpendicular (homeotropic) surface boundary conditions [13]. The thin (150 $\mu$m) or thick (1 mm) glass substrates forming cells were treated with a homeotropic polyimide SE1211 (obtained from Nissan Chemicals). The preparation of these alignment layers involved spin coating the polyimide on substrates at 2700 rpm for 30 s and then baking it for 5 min at 90 $^\circ$C, followed by additional baking for 1 h at 180 $^\circ$C. This treatment sets the strong perpendicular boundary conditions for $\bf n(r)$ at the LC-glass interface. LC cells with the gap thickness $d=5-50$ $\mu$m were produced using glass microspheres of the corresponding diameter or Mylar films of corresponding thickness interspacing the glass substrates. In some cases, wedge-shaped cells with small dihedron angles $\sim$$2^\circ$ were prepared and studied for a detailed exploration of the effects of cell thickness $d$ relative to $p$ on the soliton stability. To avoid the nematic flow-induced alignment effects, the cells were filled at an elevated temperature, right above the LC-isotropic transition, and then cooled down to the room-temperature LC phase. The 3D solitons of different types sometimes appeared during the temperature quench from isotropic phase spontaneously, but could be also generated and robustly controlled using laser tweezers when in the LC phase, as discussed below.

\subsection{Nonlinear optical imaging of preimages}

The experimental identification of 3D topological solitons relies on the analysis of preimages constructed on the basis of the nonlinear optical imaging of $\bf n(r)$ within these structures. This imaging was performed using three-photon excitation fluorescence polarizing microscopy (3PEF-PM) setup built around a BX-81 Olympus inverted optical microscope [37,38]. The polarized self-fluorescence from the LC molecules was detected within the 400-450 nm spectral range and excited through a process of three-photon absorption using a Ti-Sapphire oscillator (Chameleon Ultra II, Coherent) operating at 870 nm with 140 fs pulses at a repetition rate of 80 MHz. The 3PEF-PM signal was collected through an oil-immersion 100$\times$ objective with numerical aperture NA=1.4 and detected by a photomultiplier tube (H5784-20, Hamamatsu). We scanned the excitation beam through the sample volume with the help of galvano-mirrors (in lateral directions) and a stepper motor (across the sample thickness) and recorded the 3PEF-PM signal as a function of coordinates, which was then used to construct 3D images by means of the ParaView software (freeware obtained from KitwarePublic). The linear polarization of the excitation beam was controlled using a polarizer and a rotatable half-wave retardation plate. The detection channel utilized no polarizers. The 3PEF-PM intensity scaled as $\propto cos^6\psi$, where $\psi$ is the angle between $\bf n(r)$ and the excitation beam's linear polarization (assumed to remain unchanged despite the beam focusing through dielectric interfaces and the weakly birefringent LC medium, with the sample design minimizing these changes) and was used to reconstruct the 3D $\bf n(r)$ patterns [35, 38]. 

The reconstruction of the 3D solitonic $\bf n(r)$-structures took advantage of the self-fluorescence patterns obtained at different polarizations of excitation light, as described elsewhere [30, 35-38]. In order to eliminate the ambiguity between the two possible opposite $\bf n(r)$ tilts in the analysis of 3D images, additional cross-sectional 3PEF-PM images were obtained at orientations of the LC cell's normal tilted by $\pm 2^\circ$ with respect to the microscope axis for linear polarizations of excitation laser light parallel or perpendicular to the plane of the corresponding vertical cross-sectional image. The $\bf n(r)$ tilt ambiguity was then eliminated based on the $\propto cos^6\psi$ scaling of the 3PEF-PM signal and the ensuing spatial changes of intensity prompted by the $\pm 2^\circ$ tilts. To further narrow the angular sector of $\bf n$-orientations corresponding to preimages of points on $S^2/Z_2$ with target azimuthal angles $\phi$, we each time obtained three 3D images with azimuthal orientation of the linear polarization of excitation beam at $\phi$ and $\phi\pm3^\circ$. These 3D images were smoothed using Matlab-based software and then used in a differential analysis to improve orientational resolution to better than $\pm3^\circ$. Consistent with the nonpolar nature of $\bf n(r)$ and the ground-state manifiold of the LC [23], this 3D imaging yields preimages corresponding to a single point on $S^2/Z_2$ (without distinguishing the $\bf n$ and $-\bf n$ orientations). These experimentally reconstructed $\bf n(r)$-patterns can be vectorized by exploiting the continuity of the director field, which was done for some types of analysis that we performed. To further probe the topology of 3D solitons with nonzero Hopf indices, we then used an experimental procedure equivalent to the mathematical Hopf mapping (Fig. 1d,e) in order to relate all inter-linked closed-loop preimages in the LC sample's volume with all corresponding distinct points on $S^2/Z_2$. This new approach allowed us to experimentally probe linking of preimages within the 3D solitons and determine their Hopf indices, revealing a surprisingly large variety of topological solitons described below. The 3D nonlinear optical imaging was supplemented by conventional polarizing optical microscopy observations in the transmission mode by using the same multi-modal imaging setup built around the BX-81 Olympus inverted microscope (part of the 3PEF-PM setup described above) and a charge coupled device camera (Flea, PointGrey).

\subsection{Laser tweezers and optical generation of solitons}

To reliably control their topology, the 3D solitons were generated with optical tweezers. This optical generation utilized an Ytterbium-doped fiber laser (YLR-10-1064, IPG Photonics, operating at 1064 nm) and a phase-only spatial light modulator (P512-1064, Boulder Nonlinear Systems) integrated into a holographic laser tweezers setup capable of producing arbitrary 3D patterns of laser light intensity within the sample [13,22,30]. The laser tweezers were also integrated with the 3D imaging setup described above [30], enabling fully optical generation, control, and nondestructive imaging of the solitons. The physical mechanism behind the laser generation of solitons is the optical Fr{\'e}edericksz transition, the realignment of the LC director away from the far-field background $\bf n_0$ caused by its coupling to the optical-frequency electric field of the laser beam, which is described by a corresponding term of free energy [13], similar to that given by Eq. (3) and discussed above for the case of low-frequency electric fields. This coupling, enriched by holographically generated patterning of the trapping laser beam's intensity, phase singularities and translational motion of the individual traps [13,14,22,30,35-38], prompts complex director distortions that then relax to global or local elastic free energy minima, some of which are the topologically nontrivial solitons of interest to this study. For example, the 3D solitons with Hopf indices $Q=\pm1$ were typically laser-generated in a uniform unwound background $\bf n_0$ by moving the laser focus of the holographic optical trap along a circular trajectory within the LC cell's mid-plane. By limiting the laser power to about 50 mW and by controlling the winding direction and depth of the circular lasers beam motion, we pre-selected generation of the $Q=1$ or $Q=-1$ solitons. The elementary torons and twistions were laser-generated as described elsewhere [13,37]. Alternatively, the 5CB-based chiral nematic LCs could be locally heated to the isotropic phase of the material by a focused laser beam of power $>$200 mW, so that the spontaneous appearance of solitons could be then prompted upon quenching it back to the LC phase in a way similar to that after an initial entire-sample quench during the sample preparation. By repeating this laser-induced heating and subsequent cooling many times, we generated the desired structures despite the low probability/yield of inducing topological solitons. Although the studied 3D solitons could be generated using both of the above approaches and under multiple types of different conditions, to assure that other researchers can reproduce our results, we provide specific experimental details corresponding to all solitons that we study experimentally in the figure captions. Examples of polarizing optical micrographs of several localized 3D solitonic structures with different topologies and laser-generated next to each other are shown in Fig. 1f,g. Although the conventional polarizing optical imaging does not provide insights into the complex and beautiful topology of these structures (Fig. 1f,g), we are able to analyze it both through computer simulations and experimentally by using the 3D nonlinear optical imaging.

All solitonic structures could be further optically manipulated using the laser tweezers at powers of 2$-$5 mW, significantly lower than what is needed for their generation, without altering their topological nature. The optical gradient forces that enable this laser manipulation capability arise from the contrast of the effective refractive index between the solitons and the surrounding far-field background $\bf n_0$ [36]. With the trapping beam parallel to $\bf n_0$, the refractive index of the soliton's surrounding is the ordinary refractive index of the LC while that within the soliton is changing with coordinates between the ordinary and extraordinary values, depending on the orientation of $\bf n(r)$ (which is the LC's optical axis). The ensuing effective refractive index contrast approaches the optical refractive index anisotropy, which is $\approx$0.2 for 5CB and $\approx$0.04 for ZLI-2806 [34-38], sufficient for an effective laser manipulation [44]. Despite the fact that the solitons are just localized structures of $\bf n(r)$, without any foreign inclusions in the LC medium, their optical manipulation resembles that of particles and (at higher laser powers $>$10 mW) can be further enhanced by elastic interactions between the laser-induced director distortions and those due to the localized $\bf n(r)$-patterns of the solitons themselves [44].

\subsection{Numerical modeling approach}

Numerical modeling of the energy-minimizing $\bf n(r)$-structures was performed using a relaxation routine applied to Eqs. (1) and (2). The Frank elastic constants $K_{11}$, $K_{22}$, and $K_{33}$ that describe the energetic cost of splay, twist, and bend deformations, respectively, as well as the average constant $K$, are based on literature data for the two nematic hosts used in our study (Table 1). We assume that the surface free energy does not need to be included in the minimization problem because of the strong surface boundary conditions for n. The numerical relaxation routine calculates the spatial derivatives of $\bf n(r)$ on a computational grid using the 2nd-order finite difference scheme [36]. Commonly, the periodic boundaries are implemented along the lateral directions of the computation box while fixed perpendicular surface boundary conditions are applied at the top/bottom confining substrates to define the uniform far-field director $\bf n_0$=(0,0,1). In some of these simulations, the vertical conditions $\bf n_0$=(0,0,1) were also enforced at the lateral edges of the 3D simulation box. Both the analytical ansatz configurations [45] and the random fields were used as initial conditions in the free energy minimization, yielding similar results. At each time step $\Delta$t of the numerical simulation, we computed the functional derivatives corresponding to the Lagrange equation $\partial F/(\partial n_i ) =0$ and then also the resulting elementary displacement $\partial n_i=-\Delta t \partial F/(\partial n_i )$, where the subscript i denotes orientations along the x, y, and z axes. The maximum stable time step used in the relaxation routine is determined as $\Delta t=0.5(min(h_i ))^2/max(K)$, where $min(h_i )$ is the smallest computational grid spacing and $max(K)$ is the largest elastic constant appearing in Eqs. (1) or (2) (Table 1). The steady-state stopping condition is determined through monitoring the change of the spatially averaged functional derivative with respect to time. When this value asymptotically approaches zero, the system is assumed to be in a state corresponding to a local or global energy minimum and the relaxation procedure is terminated. The 3D spatial discretization was performed on fairly large 3D grids, such as the 112 $\times$ 112 $\times$ 32 grid. This allowed us to exclude discretization-related artifacts influencing the structural stability of solitons. 

For the grid spacing of $h_x=h_y=h_z=1$ $\mu$m and 32 grid points across the cell, the effective LC cell thickness $d=32$ $\mu$m was tuned to match that used in experiments; for LC samples of other thickness d mimicking that of experimental cells, the h-values were adjusted accordingly. In order to speed up the relaxation of energy-minimizing $\bf n(r)$-configurations corresponding to local or global energy minima, the minimization was additionally performed with a relaxation method for a two-dimensional grid of equally spaced points that was used assuming the axial symmetry and then rotated around the axial symmetry axis to obtain a volume of equally spaced voxels on a 3D grid with corresponding $\bf n(r)$-orientations. The grid spacing in this case was equal in all directions and discretized into 192 $\times$ 192 $\times$ 64 points. The free-energy-minimizing computer-simulated $\bf n(r)$-configurations obtained using different grids and discretization approaches were analyzed and compared to each other and to experiments through the generation of 3D iso-surfaces (including the preimages of single director orientations). This comparison allowed us to assure that our findings are independent of the type of grid discretization. In order to construct preimages within the 3D volume of the static topological solitons, we calculated the magnitude of the difference between a unit director (vector) defining a target point on $S^2/Z_2$ ($S^2$) and the solitonic 3D $\bf n(r)$, which was then visualized with the help of an isosurface of a small value in the ensuing scalar field that confines a 3D volume of the preimage. Combined with the experimental method of construction of preimages described above, this procedure allowed us to unambiguously assure the correspondence between numerically simulated and experimental solitonic structures. 

\end{multicols}
\begin{center} 
\noindent Table 1. Elastic constants of the nematic LC hosts and helical twisting power $\xi$ of the used chiral additives in nematic hosts.
\begin{tabular}{lcccccc}
\hline \hline
Nematic LC host  & $K_{11},~pN$&$K_{22},~pN$&$K_{33},~pN$&$K_{24},~pN$&$\xi$ of cholesterol pelargonate, $\mu m^{-1}$&$\xi$ of CB-15, $\mu m^{-1}$ \\ \hline
5CB              & ~6.4       & 3.0~        & 10.0        & 3.0~    &  -6.25   &  7.3 \\
ZLI-2806      & 14.9       & 7.9~        & 15.4        & 7.9~     &  --        &  5.9 \\
\hline \hline
\end{tabular}
\end{center}
\bigskip
\begin{multicols}{2}

\section{Results}

\subsection{Torons and hopfions with $Q=0$}

Our integrated numerical and experimental approach of imaging preimages and identification of the 3D topological solitons is analogous to the Hopf mapping from the 3D space, $\mathbb{R}^3$, to the ground-state manifold of the LC, $S^2/Z_2$, or $S^2$ for the case of vectorized $\bf n(r)$. The LC sample's spatial regions with director field $\bf n(r)$ orientations corresponding to target points on $S^2/Z_2$ (the preimages) are imaged sequentially by varying linear polarization of the 3PEF-PM excitation light, using the differential analysis to improve orientational sensitivity of this approach, and eliminating the director tilt ambiguity (see the methods section above). This nonlinear optical imaging of preimages within the solitons is based on orientation-dependent self-fluorescence of rod-like molecules of the LC that (on average) locally align with $\bf n(r)$. Consistent with the nonpolar symmetry of the LC, this 3PEF-PM based imaging approach simultaneously yields pairs of preimages corresponding to $\bf n$ and $-\bf n$, i.e. to a single point on points on $S^2/Z_2$ and to two diametrically opposite points on $S^2$ in the case when this line field is vectorized along $\bf n$ or $-\bf n$. The far-field vertical alignment of $\bf n_0$=(0,0,1) set by the strong boundary conditions at the confining surfaces, along with the continuity of $\bf n(r)$ evidenced by absence of singular light-scattering defects, provide the foundations for analyzing $\bf n(r)$ structures based on the strong $\propto cos^6\psi$ orientational dependence of the 3PEF-PM signal. The numerical analogs of the experimentally reconstructed preimages are obtained from analyzing the 3D director patterns that minimize the elastic free energy, as described above.

The simplest observed 3D solitonic structure is shown in Fig. 2. The in-plane and vertical cross-sections of its spatially localized $\bf n(r)$ are presented with the help of double-cones, which are colored according to two different schemes (Fig. 2a-d), both designed to be consistent with the nonpolar nature $n\equiv-n$ of the director field. The coloring scheme that show the two cones within each double-cone in different colors corresponding to that of respective diametrically opposite points on $S^2$ also allows one to analyze the two mutually opposite vector fields vectorizing $\bf n(r)$ (Fig. 2a,c). The 

\end{multicols}
\begin{figure}[H]
\begin{center}
\includegraphics[width=0.6\textwidth]{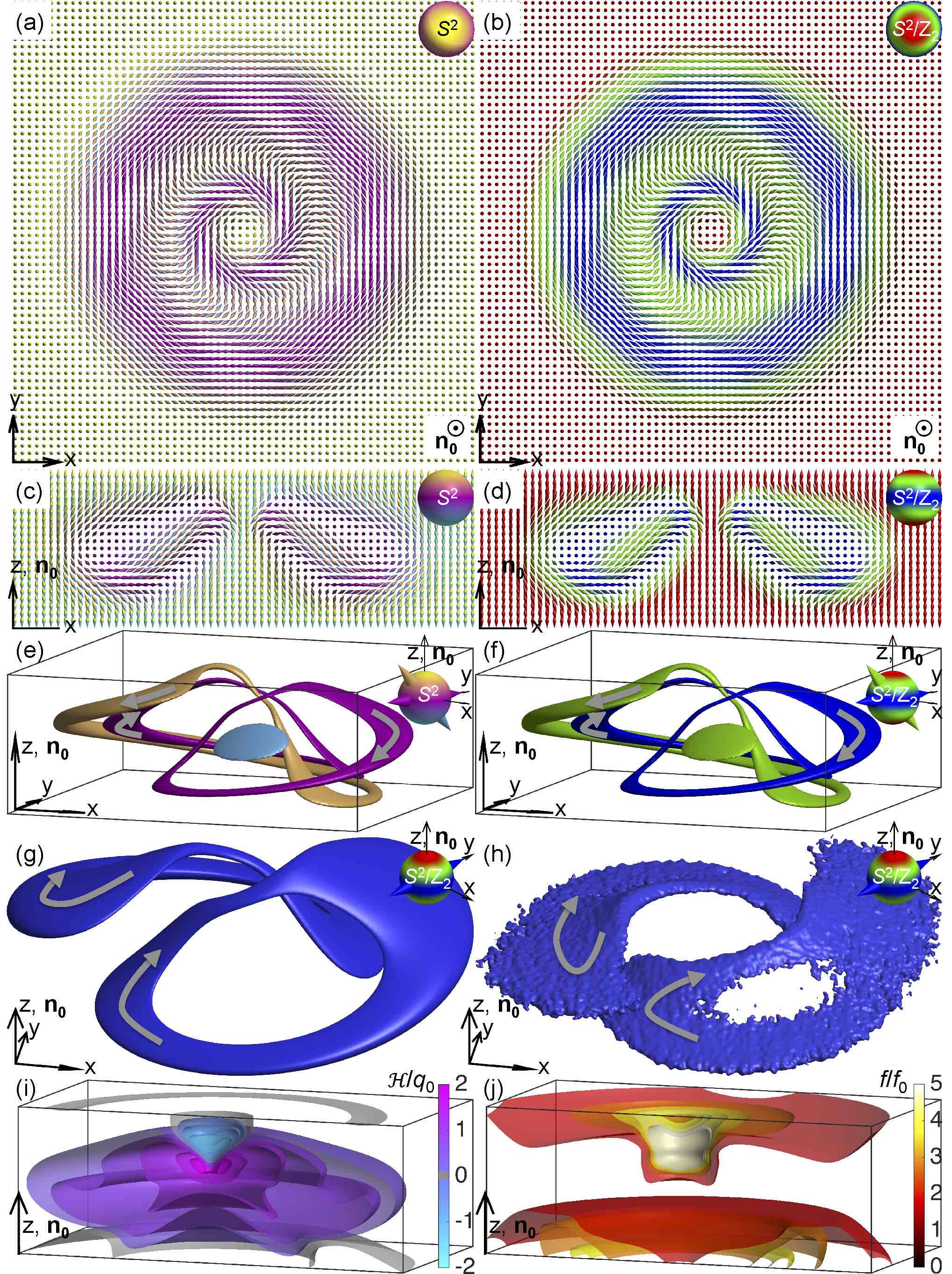}
\end{center}
\caption{3D solitons with $Q=0$. (a-d) Computer simulated (a,b) in-plane and (c,d) vertical cross-sections of $\bf n(r)$ depicted using double cones and two different color schemes that establish correspondence between director orientations and the points on $S^2/Z_2$ (top-right insets). (e,f) Computer-simulated preimages of a $Q=0$ 3D soliton for two sets of the diametrically opposite points on $S^2/Z_2$ marked by double cones in the top-right insets for two different coloring schemes of the $S^2/Z_2$. (g,h) Comparison of representative (g) computer-simulated and (h) experimental preimages of the $Q=0$ soliton for two diametrically opposite points on the ``equator'' of $S^2/Z_2$ (top-right insets). Gray arrows in (e-h) indicate the consistently determined circulations of the preimages. (i,j) Perspective views of the 3D computer-simulated isosurfaces of normalized (i) handedness and (j) free energy density for this axially symmetric solitonic structure, with the used color schemes provided in the right-side insets. The experimental preimages were reconstructed based on 3PEF-PM images obtained for structures in a 5CB-based partially polymerizable nematic LC mixture with chiral additive CB-15, $d =10\mu$m and $d/p\sim1$; before imaging, the unpolymerized 5CB was replaced by immersion oil. Computer simulations were preformed for 5CB elastic constants and assuming $K_{24} = 0$.}
\end{figure}
\begin{multicols}{2}

\noindent nonsingular axially symmetric structure of this soliton introduces a spatially localized twisted region embedded in the uniform far-field $\bf n_0$. Mapping the vectorized $\bf n(r)$ from the soliton's cross-section (Fig. 2c) onto $S^2$ does not fully cover it, albeit the similar mapping of nonpolar $\bf n(r)$ onto $S^2/Z_2$ covers this ground-state manifold fully, with some parts of $S^2/Z_2$ covered twice (Fig. 2d). A similar analysis can be done for preimages of vectorized and nonpolar $\bf n(r)$ in the entire 3D volume of the soliton. The vectorized $\bf n(r)$-pattern of these solitons has closed-loop preimages for a majority of points on $S^2$ (see examples in Fig. 2e), except for the vicinity of its south pole. In the case of the nonpolar $\bf n(r)$ and the $S^2/Z_2$ ground-state manifold, the preimages of most of the points are pairs of unlinked closed loops (Fig. 2f), albeit one of the two loops shrinks into a disc and disappears for points near the north/south pole of $S^2/Z_2$. The computer-simulated preimages closely match their experimental counterparts, as shown by comparing examples of two-loop preimages of the same single point (marked by the blue double cone) on $S^2/Z_2$ in Fig. 2g,h, with the consistently chosen circulation directions marked using curved gray arrows. The numerical energy-minimizing $\bf n(r)$-configuration allows us to plot the handedness of the director twist [37,46], defined as $\mathcal{H}= -\bf n \cdot (\nabla \times \bf n)$ and normalized by $q_0$, to observe that $\mathcal{H}$ matches the intrinsic chiral nematic LC's handedness within most of the volume of the soliton, except for a small region close to to one of the confining substrates, where it reverses and becomes opposite (Fig. 2i). This observation hints that such a 3D soliton is stabilized by the chiral LC medium's tendency to twist, which is confirmed by a 3D plot of free energy density isosurfaces of the soliton (Fig. 2j). Importantly, the free energy within the soliton is below that of the uniform unwound state almost everywhere except for the small localized region matching the region of $\mathcal{H}$ opposite to that of LC's chirality (Fig. 2i,j). 

The nonsingular 3D soliton shown in Fig. 2 is rather different from the simplest toron structure that we introduced in Ref. [13], with $\pi$-twist from its central axist to the periphery in all radial directions (Fig. 3). The midplane cross-section of the toron is a two-dimensional skyrmion (baby skyrmion), which belongs to the second homotopy group $\pi_2(S^2)=\mathbb{Z}$ [30]. Mapping the vectorized director from the elementary toron's midplane (Fig. 3a) onto the $S^2$ order parameter space covers the sphere once, indicating that the skyrmion number is equal unity [30]. In the overall 3D configuration of the elementary toron, the localized twisted region is capped with two hyperbolic point defects near confining surfaces, as we discussed in detail in Ref. [13]. The preimages of this simplest toron have shapes of bands (two per single point on $S^2/Z_2$ and a single band in the case of a vectorized field and the $S^2$ ground-state manifold) terminating on the point defects. Interrupted by the two point singularities, the two-band preimages of points on $S^2/Z_2$ of these torons do not form closed loops (Fig. 3c-e), albeit this behavior is very different for the new types of 

\end{multicols}
\begin{figure}[H]
\begin{center}
\includegraphics[width=0.6\textwidth]{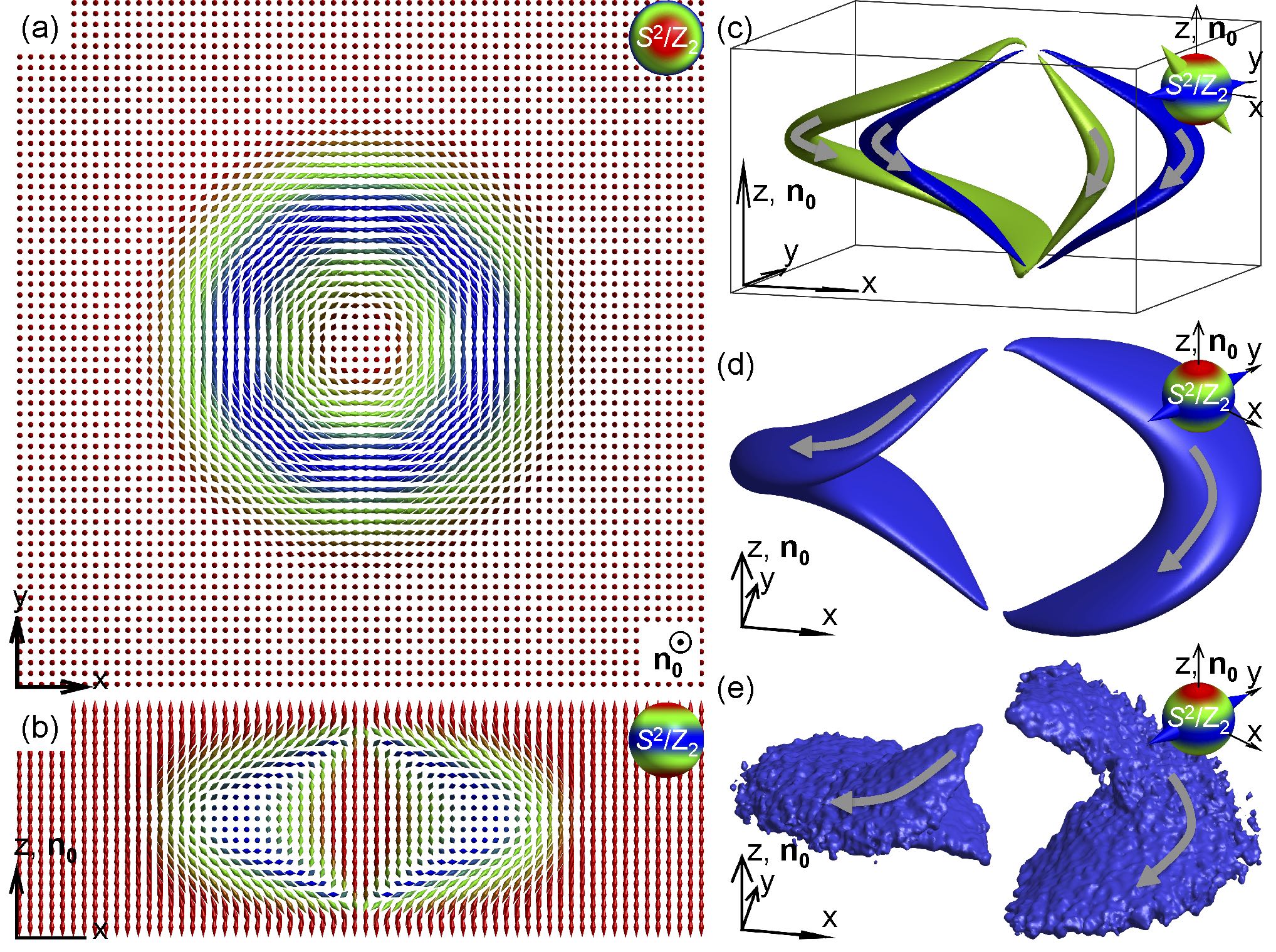}
\end{center}
\caption{Elementary toron structure in a confined chiral nematic LC. (a,b) Computer simulated (a) in-plane and (b) vertical cross-sections of $\bf n(r)$ of the toron depicted using double cones and the color scheme that establishes correspondence between director orientations and the points on $S^2/Z_2$ (top-right insets). (c) Computer-simulated preimages of the toron for two sets of the diametrically opposite points on $S^2/Z_2$ marked by double cones in the top-right inset. (d,e) Comparison of representative (d) computer-simulated and (e) experimental preimages of the toron for two diametrically opposite points on the ``equator'' of $S^2/Z_2$ (top-right insets). The experimental preimages in (e) were reconstructed based on 3PEF-PM images of structures in a 5CB-based partially polymerizable nematic LC mixture with chiral additive CB-15, $d =10\mu$m and $d/p\sim1$; before imaging, the unpolymerized 5CB was washed out and replaced by immersion oil. Gray arrows indicate the consistently determined circulations of the preimages. Computer simulations were preformed for 5CB elastic constants while also assuming $K_{24} = 0$.}
\end{figure}
\begin{multicols}{2}

\noindent torons with larger amounts of twist and nontrivial closed-loop preimages that we discuss below. The torons, ensembles of the nonsingular solitonic twisted structures and singular point defect or disclination loop entities, can be classified not only by the types of the self-compensating singular defects [13], but also by the linking of preimages of their solitonic parts [36], as we discuss in detail below for the cases of new, complex toron structures.

\subsection{Elementary hopfions with $Q=\pm1$ and linking of preimages}

The solitons shown in Figs. 4 and 5 also embed the axially symmetric twisted regions into the uniform background $\bf n_0$, but their solitonic $\bf n(r)$ twists by 2$\pi$ in all radial directions from the 3D soliton's central axis (parallel to $\bf n_0$) to the far-field periphery (Fig. 4a,b and Fig. 5a,b). For both of these  

\end{multicols}
\begin{figure}[H]
\begin{center}
\includegraphics[width=0.8\textwidth]{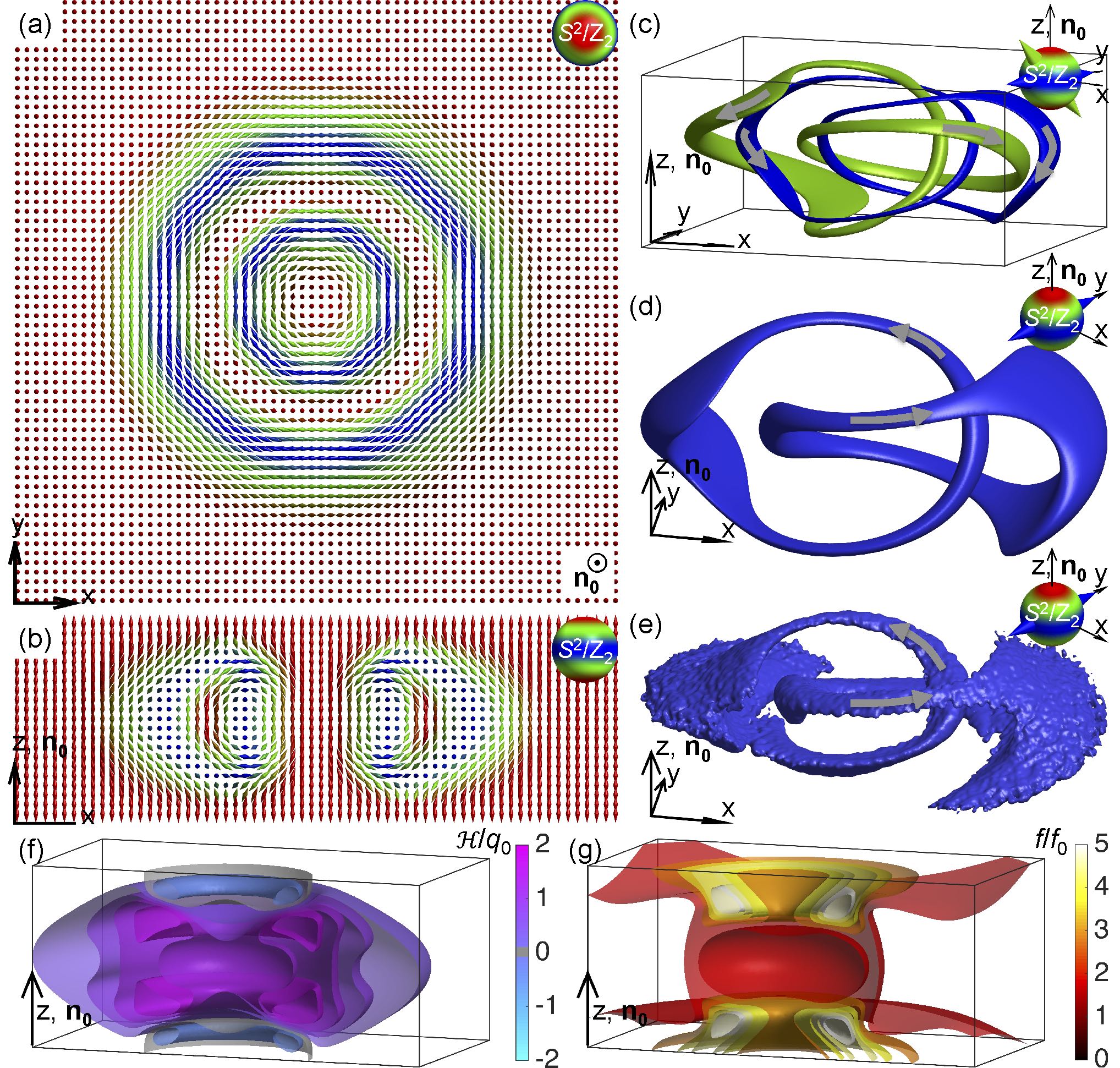}
\end{center}
\caption{3D topological soliton with $Q=1$. (a,b) Computer simulated (a) in-plane and (b) vertical cross-sections of the axially symmetric $\bf n(r)$-structure of the hopfion depicted using double cones and the color scheme that establishes correspondence between director orientations and the points on $S^2/Z_2$ (top-right insets). (c) Computer-simulated preimages of the hopfion for two sets of the diametrically opposite points on $S^2/Z_2$ marked by double cones in the top-right inset. (d,e) Comparison of representative (d) computer-simulated and (e) experimental preimages of the hopfion for two diametrically opposite points on the ``equator'' of $S^2/Z_2$ (top-right insets). Computer simulations were preformed for elastic constants of 5CB while also assuming $K_{24}=0$. The preimages were reconstructed based on 3PEF-PM images of structures in a 5CB-based partially polymerizable nematic LC mixture with a chiral additive CB-15, in a cell with $d =10\mu$m and $d/p\sim1$; before imaging, the unpolymerized 5CB was replaced by immersion oil. Linking of the two closed loops establishes the Hopf index $Q=1$, as discussed in the text. Gray arrows in (c-e) indicate the consistently determined circulations of the preimages. (f,g) Perspective views of the 3D computer-simulated isosurfaces of normalized (f) handedness and (g) free energy density for this axially symmetric hopfion, with the corresponding color schemes provided in the right-side insets.}
\end{figure}
\begin{multicols}{2}

\end{multicols}
\begin{figure}[H]
\begin{center}
\includegraphics[width=0.8\textwidth]{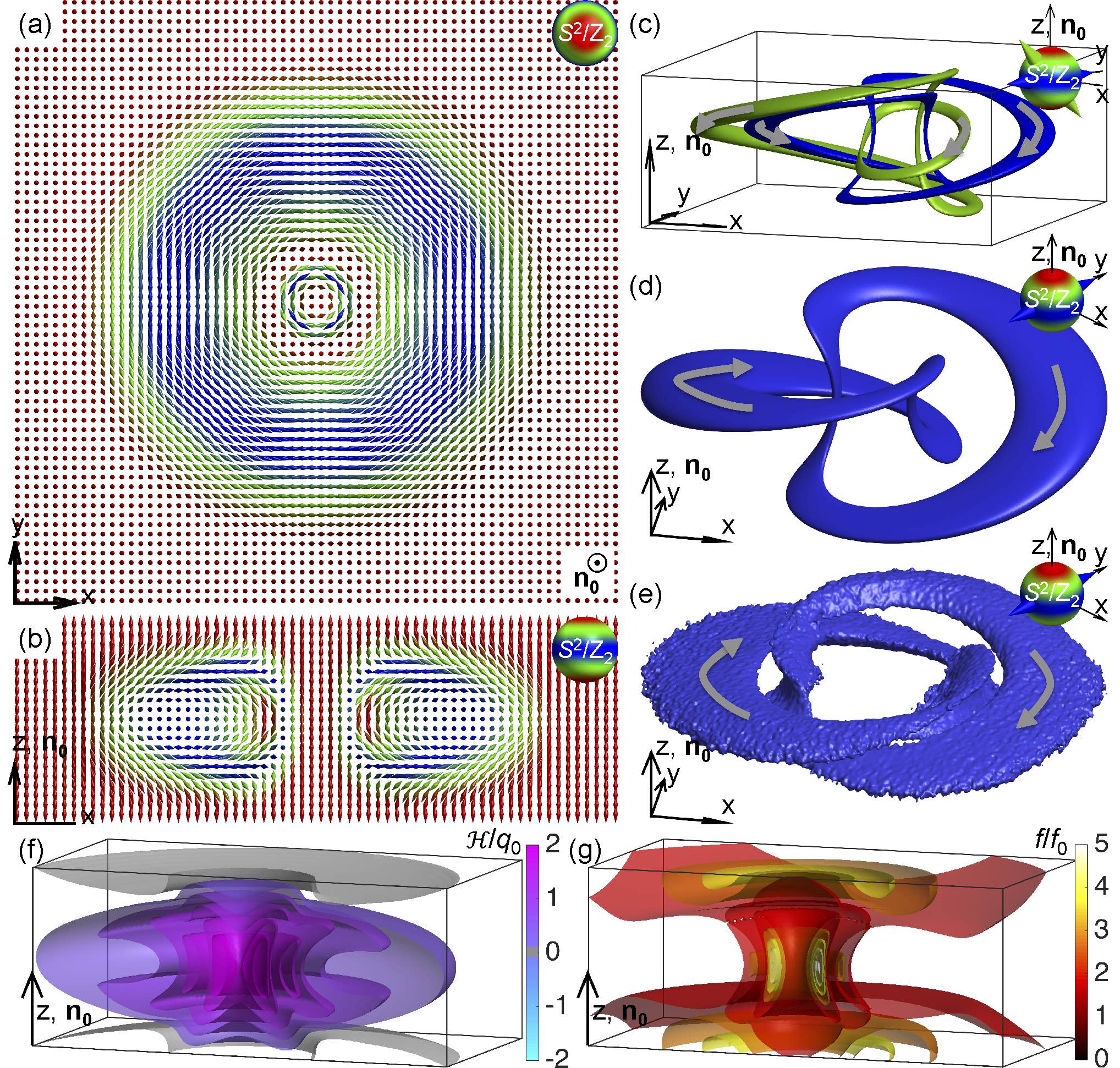}
\end{center}
\caption{3D topological soliton with $Q=-1$. (a,b) Computer simulated (a) in-plane and (b) vertical cross-sections of the axially symmetric $\bf n(r)$-structure of the $Q=-1$ hopfion depicted using double cones and the color scheme that establishes correspondence between director orientations and the points on $S^2/Z_2$ (top-right insets). (c) Computer-simulated preimages of the hopfion for two sets of the diametrically opposite points on $S^2/Z_2$ marked by double cones in the top-right inset. (d,e) Comparison of representative (d) computer-simulated and (e) experimental preimages of the hopfion for two diametrically opposite points on the ``equator'' of $S^2/Z_2$ (top-right insets). Linking of the two closed loops establishes the Hopf index $Q=-1$, as discussed in the text. Computer simulations were preformed for elastic constants of 5CB while also assuming $K_{24}=0$. The preimages were reconstructed based on 3PEF-PM images of structures in a 5CB-based partially polymerizable nematic LC mixture with a chiral additive CB-15, in a cell with $d =10\mu$m and $d/p\sim1$; before imaging, the unpolymerized 5CB was replaced by immersion oil. Gray arrows in (c-e) indicate the consistently determined circulations of the preimages. (f,g) Perspective views of the 3D computer-simulated isosurfaces of normalized (f) handedness and (g) free energy density for this axially symmetric hopfion, with the corresponding color schemes shown in the right-side insets.}
\end{figure}
\begin{multicols}{2}

\noindent solitons, all points on $S^2/Z_2$ have preimages in the form of two inter-linked closed loops (Fig. 4c-e and 5c-e). Experimental closed-loop preimages closely match their theoretical counterparts (Figs. 4d,e and 5d,e) and all wind around each other to link once. In the case of a vectorized $\bf n(r)$, for both solitons, the preimages of every single point on $S^2$ are single closed loops and such preimages of any two distinct points on $S^2$ are linked once. As discussed above, the Hopf index of such a soliton $Q=\sum C/2$ (determined by analyzing the crossings C) is the linking numbers of preimages for vectorized $\bf n(r)$ or, equivalently, the linking number of the two closed loops that form a single preimage of any point on $S^2/Z_2$ [31]. By vectorizing $\bf n(r)$ of the two types of observed solitons (Figs. 4 and 5), so that the $\bf n_0$ points in the same directions for both of them and so that the corresponding circulations of the far-field preimages define circulations of all other preimages through the requirement of continuity, we find the opposite $Q=\pm1$ Hopf indices characterizing the two solitons. Hopf links of the two closed loops that constitute preimages of all points on $S^2/Z_2$ for these two different solitons in the case of nonpolar $\bf n(r)$, the corresponding Hopf links of single-loop preimages of any two points on $S^2$ for vectorized $\bf n(r)$ and the LC in which these solitons are hosted are all chiral in nature. By using a vectorized $\bf n(r)$ and choosing the circulation of the preimage of the north pole on $S^2$ to be along $\bf n_0$ through the soliton's center, we consistently define circulations of all other preimages while smoothly exploring $S^2$. We find that thus determined linking number stays conserved for all pairs of vectorized-$\bf n(r)$ preimages of $S^2$-points within the same solition, yielding its Hopf index, which is $Q=1$ for the soliton shown in Fig. 4 and $Q=-1$ for the soliton depicted in Fig. 5. Interestingly, the $Q$-values stay the same upon inverting the vectorization direction $\bf n(r) \rightarrow -\bf n(r)$, which is different from the case of hedgehog charges of point defects in LCs [47] that change sign in response to the $\bf n(r) \rightarrow -\bf n(r)$ operation. However, taking a mirror image negates the linking numbers of all Hopf links and the corresponding $Q$ values while also transforming a left-handed LC host into its right-handed counterpart, which is again different from the hedgehog charges of point defects that would stay unchanged during this operation [14,47]. These $Q=\pm1$ solitons are topologically similar to their counterpart in colloidal fluid chiral ferromagnets [22] that we observed recently (to be reported elsewhere), albeit here they are realized in a non-polar line field rather than in a vector field. Because of the nonzero Hopf index values, similar to the case of their ferromagnetic counterparts, we will refer to these solitons as ``hopfions''. Any changes of their quantized Hopf index require destroying the orientational order or generating singular defects and, thus, overcoming free energy barriers associated with them. This helps to stabilize such topological solitons with $Q\neq 0$ as metastable or stable field configurations, albeit the main physical mechanism responsible for their stability is the same as for the solitons with $Q=0$ (Figs. 2 and 3) and is related to the LC medium's tendency to twist $\bf n(r)$ with the spatial periodicity comparable to its intrinsic helicoidal pitch $p$. Prevailing parts of the soliton's volume have twist handedness matching that of the LC hosting them (Figs. 4f and 5f), although there are some small regions within the solitons with $\mathcal{H}$ opposite to that of the LC's intrinsic chirality. The 3D isosurfaces of free energy density convincingly show that such a 3D topological soliton embeds a large region of low-energy twisted $\bf n(r)$ (lower than that of the surrounding unwound uniform background) and that its stability is helped by the medium's chirality (Figs. 4g and 5g). 

\subsection{3D solitons with multi-loop preimages}

The elementary $Q=\pm1$ hopfions are not the only 3D solitons with linking of closed-loop preimages. Many interesting topologically nontrivial configurations arise with the increase of the amount of $\bf n(r)$-twist embedded in the uniform unwound background $\bf n_0$. For example, $\bf n(r)$ twists by 4$\pi$ in all radial directions from the localized configuration's central axis (parallel $\bf n_0$) to the periphery of the 3D solitons shown in Figs. 6, 7 and 8. These axially symmetric solitons (Figs. 6a,b,j,k, 7a,b and 8a,b) have two-closed-loops preimages of all $S^2$ points in the case of a vectorized $\bf n(r)$ while the preimages of each point on $S^2/Z_2$ of the nonpolar director field comprise four individual closed loops (Figs. 6d,e,m,n, 7c,d and 8c,d). The preimages of $S^2$-points for the same polar angle $\theta$ of vectorized $\bf n(r)$ with respect to $\bf n_0$ but corresponding to its different azimuthal orientations tile into tori surfaces (Figs. 6f,o, 7e and 8e). There are always two such tori for a given $\theta$ (Figs. 6f,o, 7e and 8e), which is different from the case of elementary hopfions (Figs. 4 and 5), for which there is only one torus surface for each $\theta$-value. Although all four examples of solitons shown in Figs. 6-8 have preimages in the form of two separate closed loops for every single point on $S^2$ for vectorized $\bf n(r)$, the nature and topology of inter-linking of these closed-loop preimages is different, as we discuss below. 

For all $S^2$-points in the case of the vectorized $\bf n(r)$ of the soliton shown in Fig. 6a-i, the individual preiamges are formed by two separate unlinked closed loops while preimages of two separate $S^2$-points form two Hopf links with the linking number +1 for each of them (Fig. 6d,e). The sloitons shown in Fig. 6j-r also have preimages comprised of two separate unlinked closed loops and preimages of two separate $S^2$-points forming two Hopf links, but the linking number characterizing these two individual Hopf links is $-1$ (Fig. 6m,n). For all constant values of the polar angle $\theta$ characterizing orientation of $\bf n(r)$, the series of preimages with different azimuthal orientations of $\bf n(r)$ tile into two separate tori surfaces (Fig. 6f,o). By scanning the $\bf n(r)$ orientations from $\bf n(r)$=(0,0,-1) to the far $\bf n(r)$=(0,0,1)=$\bf n_0$, we find that the two tori formed by preimages of constant $\theta$ remain separate until merging with the far field background when $\bf n(r)$ becomes parallel to $\bf n_0$ (Fig. 6g,p). The behavior of the individual $S^2/Z_2$-preimages corresponding to nonpolar $\bf n(r)$ is reminiscent to that of pairs of preimages of $S^2$-points for vectorized $\bf n(r)$ (Fig. 6d-g and m-p). The preimages of $S^2$-points in the vicinity of the north pole are two separate tori that characterize $\bf n(r)$-orientations smoothly transforming to $\bf n_0$, with the far-field preimages connected through infinity (Fig. 6g,p). Thus, one can interpret the two solitonic structures shown in Fig. 6 as being formed by a coaxial arrangement of two separate hopfions of $Q=1$ (Fig. 6a-i) and $Q=-1$ (Fig. 6j-r). In addition to these two examples, we have observed other variations of such solitons that could be thought of as being formed by elementary hopfions of opposite signs of Q, including the ones with a $Q=1$ hopfion in the interior and $Q=-1$ hopfion in the exterior of the coaxial hybrid solitons and vice versa. 

\end{multicols}
\begin{figure}[H]
\begin{center}
\includegraphics[width=0.6\textwidth]{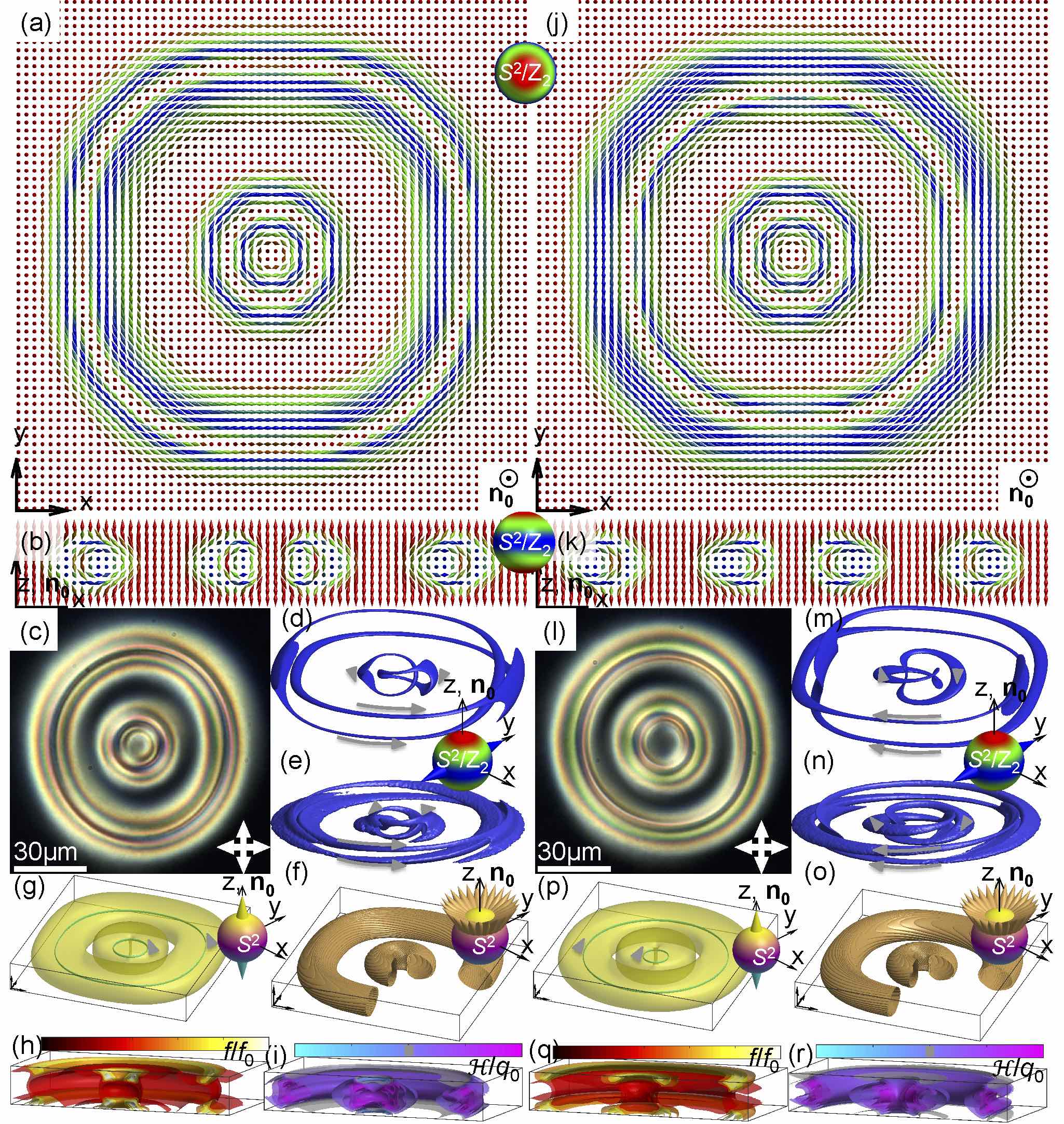}
\end{center}
\caption{3D topological solitons formed by coaxial arrangement of two different elementary hopfions of the same signs $Q=\pm1$. (a,b) Computer simulated (a) in-plane orthogonal to $\bf n_0$ and (b) vertical containing $\bf n_0$ cross-sections of the axially symmetric $\bf n(r)$-structure of a soliton with two-loop preimages depicted using colored double cones; the color scheme that establishes the correspondence between director orientations and the points on $S^2/Z_2$ (top-right insets). (c) A polarizing optical micrograph of such a 3D soliton in a confined chiral nematic LC; white double arrows show crossed polarizers. (d) Computer-simulated and (e) experimental preimages of the soliton for the diametrically opposite points on $S^2/Z_2$ marked by double cones in the right-side inset. By analyzing such preimages of all points on $S^2/Z_2$, we find no inter-linking between the preimages of the two separate hopfions and their Hopf indices $Q=1$. (f) For a constant polar angle value (inset), the closed-loop preimages of individual points on $S^2$ tile into two separate tori surfaces sharing the same vertical axis parallel to $\bf n_0$. (g) preimages of the north and south poles of $S^2$ for the vectorized director field. (h,i) Perspective views of the 3D computer-simulated isosurfaces of normalized (h) handedness and (i) free energy density for this axially symmetric soliton, with the corresponding color schemes shown in the insets above them. (j,k) Computer simulated (j) in-plane and (k) vertical cross-sections of an axially symmetric $\bf n(r)$-structure of a soliton with two-loop preimages depicted using colored double cones and the same color scheme as in (a,b) that establishes the correspondence between director orientations and the $S^2/Z_2$-points. (l) A polarizing optical micrograph of such a 3D soliton. (m) Computer-simulated and (n) experimental preimages of the hopfion for the diametrically opposite points on $S^2/Z_2$ marked by double cones in the right-side inset. (o) For a constant polar angle value, the closed-loop preimages of the individual points on $S^2$ tile into two separate tori that share the vertical axis parallel to $\bf n_0$. By analizing preimages of all points on $S^2/Z_2$, we find no inter-linking between the preimages of the two separate hopfions and also their Hopf indices $Q=-1$. (p) preimages of the north and south poles of $S^2$ for the vectorized director field. (q,r) Perspective views of the 3D computer-simulated isosurfaces of normalized (q) handedness and (r) free energy density for this soliton, with the corresponding color schemes shown in the insets. Computer simulations were preformed for elastic constants of 5CB while also assuming $K_{24}=0$. The polarizing optical micrographs in (c) and (l) were obtained for structures in a 5CB-based partially polymerizable nematic LC mixture with a chiral additive CB-15 in a cell with thickness $d =10\mu$m and $d/p\sim1$. Gray arrows in (d,e,m,n,g,p) indicate the consistently determined circulations of the preimages. The 3D preimages were reconstructed based on 3PEF-PM images of these structures after the unpolymerized 5CB was replaced by immersion oil.}
\end{figure}
\begin{multicols}{2}

A series of other solitons with 4$\pi$ twist in radial directions have a very different linking of preimages (Figs. 7 and 8 and supplementary videos S1 and S2). To analyze them, we first observe that the so-called ``Pontryagin-Thom constructions'' [14], isosurfaces of zero $z$-component of the director, $n_z=0$ (corresponding to $\theta=\pi/2$), colored by the azimuthal orientation of the in-plane $\bf n(r)$ (Figs. 7f-h and 8f-h), 

\end{multicols}
\begin{figure}[H]
\begin{center}
\includegraphics[width=0.6\textwidth]{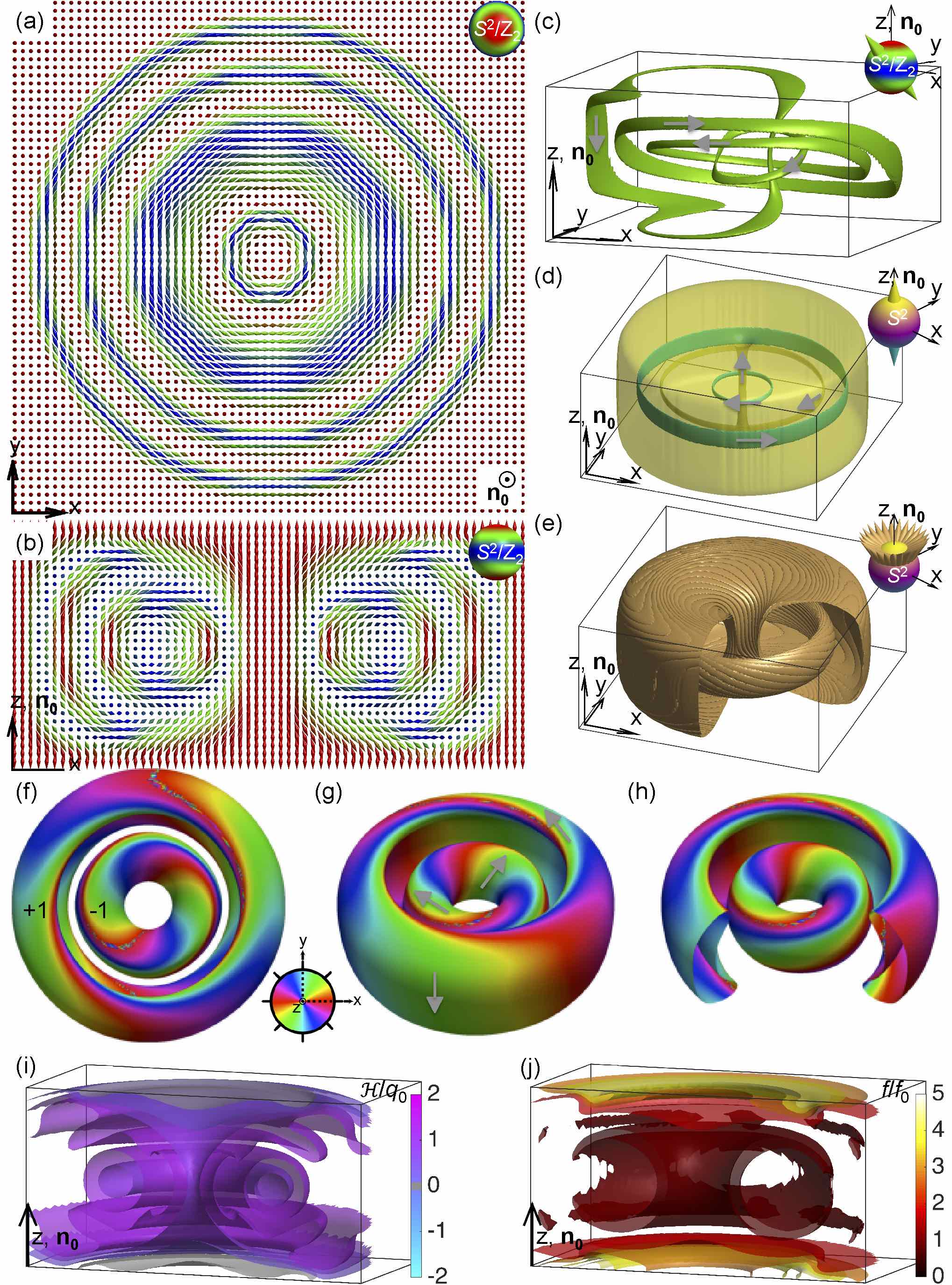}
\end{center}
\caption{3D soliton with complex linking and 4$\pi$ twist from its central axis to the far-field periphery. (a,b) Computer simulated (a) in-plane and (b) vertical cross-sections of the axially symmetric $\bf n(r)$-structure of the 3D soliton depicted using double cones and the color scheme that establishes correspondence between director orientations and the points on $S^2/Z_2$ (top-right insets). (c) A representative computer-simulated preimage of the hopfion for the diametrically opposite points on $S^2/Z_2$ marked by double cones in the top-right inset. The preimage is comprised of four inter-linked closed loops. (d,e) Computer-simulated preimages of the 3D soliton in a vectorized $\bf n(r)$ (d) for two diametrically opposite points on $S^2$ corresponding to its north and south poles (top-right inset) and (e) for a set of points characterized by a constant polar angle $\theta$ and forming a circle parallel to the spheres equator (top-right inset). Note that the preimages in (e) reside on two nested tori surfaces. A large variety of preimages of this soliton are shown in the supplementary video S1. (f-h) Three representative views on the isosurfaces of $\theta=\pi/2$ ($n_z=0$) colored by azimuthal orientations of $\bf n(r)$ according to the scheme shown in the right-side inset of (f). The numbers on top of the tori shown in (f) indicate the linking numbers that characterize the inter-linking of colored closed-loop bands and preimages of $\bf n(r)$ corresponding to points on the equator of $S^2/Z_2$. (i,j) Perspective views of the 3D computer-simulated isosurfaces of normalized (i) handedness and (j) free energy density for this axially symmetric 3D soliton, with the corresponding color schemes shown in the right-side insets. Gray arrows in (c,d) and on the green bands of (g) indicate the consistently determined circulations of the preimages. Computer simulations were preformed for elastic constants of ZLI-2806 (Table 1) and $d/p=2$.}
\end{figure}
\begin{multicols}{2}

\end{multicols}
\begin{figure}[H]
\begin{center}
\includegraphics[width=0.6\textwidth]{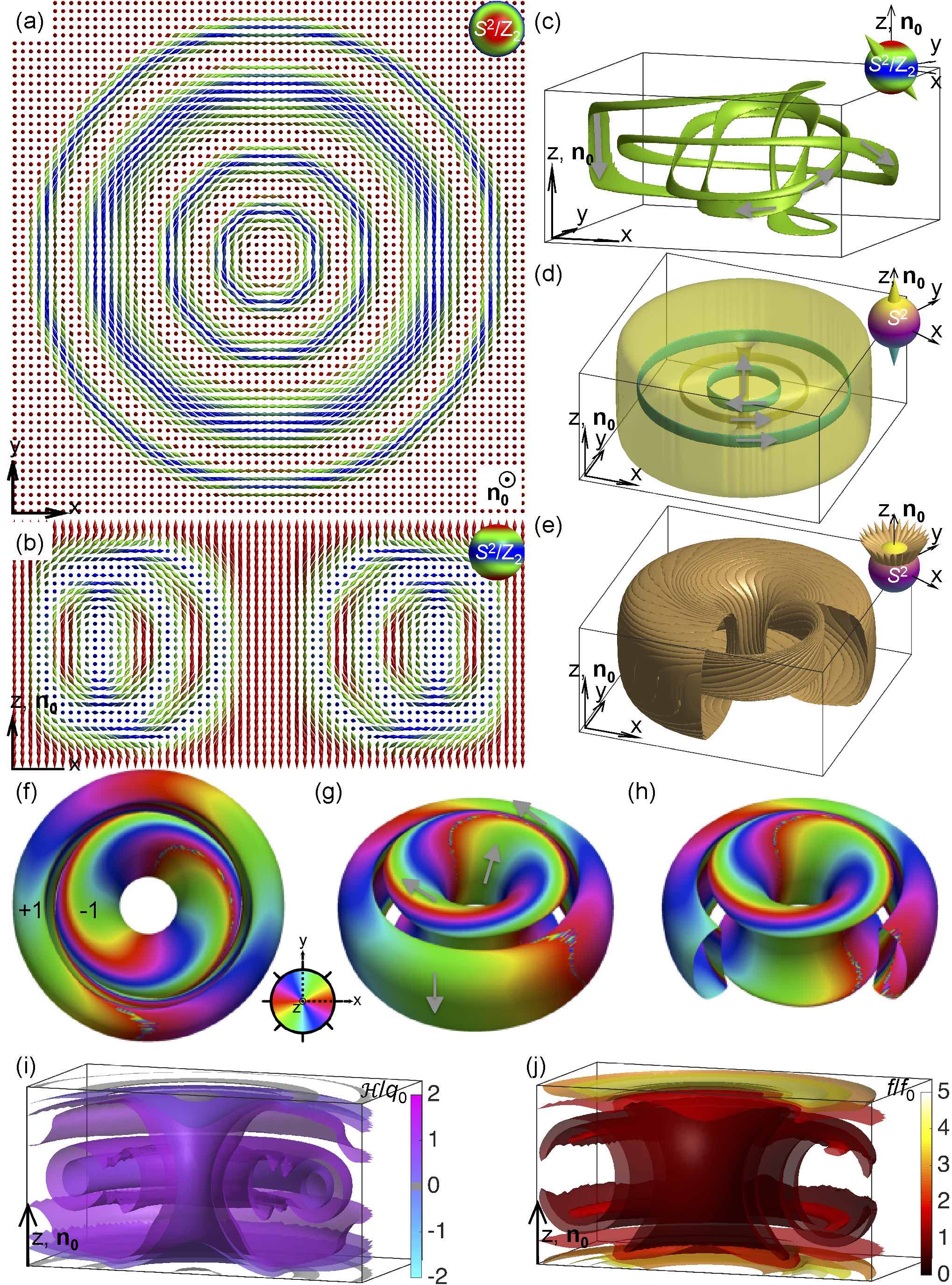}
\end{center}
\caption{3D soliton with complex linking and 4$\pi$ twist from its central axis to the far-field periphery. (a,b) Computer simulated (a) in-plane and (b) vertical cross-sections of the axially symmetric $\bf n(r)$-structure of the 3D soliton depicted using double cones and the color scheme that establishes correspondence between director orientations and the points on $S^2/Z_2$ (top-right insets). (c) A representative computer-simulated preimage of the hopfion for the diametrically opposite points on $S^2/Z_2$ marked by double cones in the top-right inset. The preimage is comprised of four inter-linked closed loops. (d,e) Computer-simulated preimages of the 3D soliton in a vectorized $\bf n(r)$ (d) for two diametrically opposite points on $S^2$ corresponding to its north and south poles (top-right inset) and (e) for a set of points characterized by a constant polar angle $\theta$ and forming a circle parallel to the spheres equator (top-right inset). Note that the preimages in (e) reside on two nested tori surfaces. A large variety of preimages of this soliton are shown in the supplementary video $S^2$.  (f-h) Three representative views on the isosurfaces of $\theta=\pi/2$ ($n_z=0$) colored by azimuthal orientations of $\bf n(r)$ according to the scheme shown in the right-side inset of (f). The numbers on top of the tori shown in (f) indicate the linking numbers that characterize the inter-linking of colored closed-loop bands and preimages of $\bf n(r)$ corresponding to points on the equator of $S^2/Z_2$. (i,j) Perspective views of the 3D computer-simulated isosurfaces of normalized (i) handedness and (j) free energy density for this axially symmetric 3D soliton, with the corresponding color schemes shown in the right-side insets. Gray arrows in (c,d) and on the green bands of (g) indicate the consistently determined circulations of the preimages. Computer simulations were preformed for elastic constants of ZLI-2806 and $d/p=2$.}
\end{figure}
\begin{multicols}{2}

\noindent also form two separate tori. The like-colored closed-loop bands of constant azimuthal $\bf n(r)$-orientation on these surfaces of two separate tori link with each other once, with the consistently determined circulation directions shown with the gray curved arrows on the green bands. These colored bands covering the $\theta=\pi/2$ tori surfaces represent all preimages of points on the equator of the $S^2$ (or $S^2/Z_2$) and are shown using the color scheme chosen to be consistent with the nonpolar nature of $\bf n(r)$ (Figs. 7f-h and 8f-h) (note that the $\bf n$ and $-\bf n$ bands are shown using the same color). The linking number of each pair of like-colored closed-loop bands is $\pm$1, opposite for the two tori comprising each of the solitons (Figs. 7f-h and 8f-h). By analyzing only the Pontryagin-Thom constructions in Figs. 7f-h and 8f-h, one could assume that the two structures are homeomorphic to each other, despite the structural differences seen in the cross-sections in Fig. 7b and 8b. However, we will show below that the two solitons are topologically distinct from each other. Interestingly, the $\theta=\pi/2$ preimages shown as closed loops colored according to azimuthal orientations of $\bf n(r)$ link differently from preimages of the north-pole point on $S^2$ and the $S^2$-points in its vicinity (Figs. 7c-e and 8c-e). Moreover, the two-tori surfaces of constant $\theta$ are nested one in another for $\theta<\theta_c$, but become separate from each other within $\theta_c<\theta<180^\circ$, where the critical polar angle is $\theta_c \approx 74^\circ$ for the soliton shown in Fig. 7 and $\theta_c \approx$87$^\circ$ for that in Fig. 8. Most interestingly, such transformation of the two-tori surfaces upon changing $\theta$ takes place without compromising the soliton's nonsingular nature and the pre-images on the different tori align with each other during the re-linking (supplementary videos S1 and S2). For the vectorized $\bf n(r)$, the two-loop preimages of single $S^2$-points are linked once at $\theta<\theta_c$ but unlinked at $\theta>\theta_c$. The linking of preimages of different $S^2$-points of such 3D solitons depends on the locations of these points on $S^2$ (Figs. 7c-h and 8c-h) and is not a conserved quantity. The nature of linking of four-loop preimages of distinct points on $S^2/Z_2$ for these solitons in a nonpolar $\bf n(r)$ is even more complex. To characterize it, we use simplified-topology and graph presentations (Table 2) of the closed-loop preimages and their linking. In these graphs, the closed-loop components of preimages are shown as filled circles colored according to the positions of corresponding points on the ground-state manifold and the individual links are indicated by black edges connecting the corresponding circles. This presentation allows us to provide an exhaustive set of possibilities for inter-linking of preimages of two distinct points on the $S^2/Z_2$ or $S^2$ depending on their relative locations (Table 2). Moreover, the summary of the preimage linking in Table 2 reveals differences in topology of the two solitons shown in Figs. 7 and 8, which is manifested by the differences in consistently defined preimage circulations.

\end{multicols}
\begin{center}
\includegraphics[width=1\textwidth]{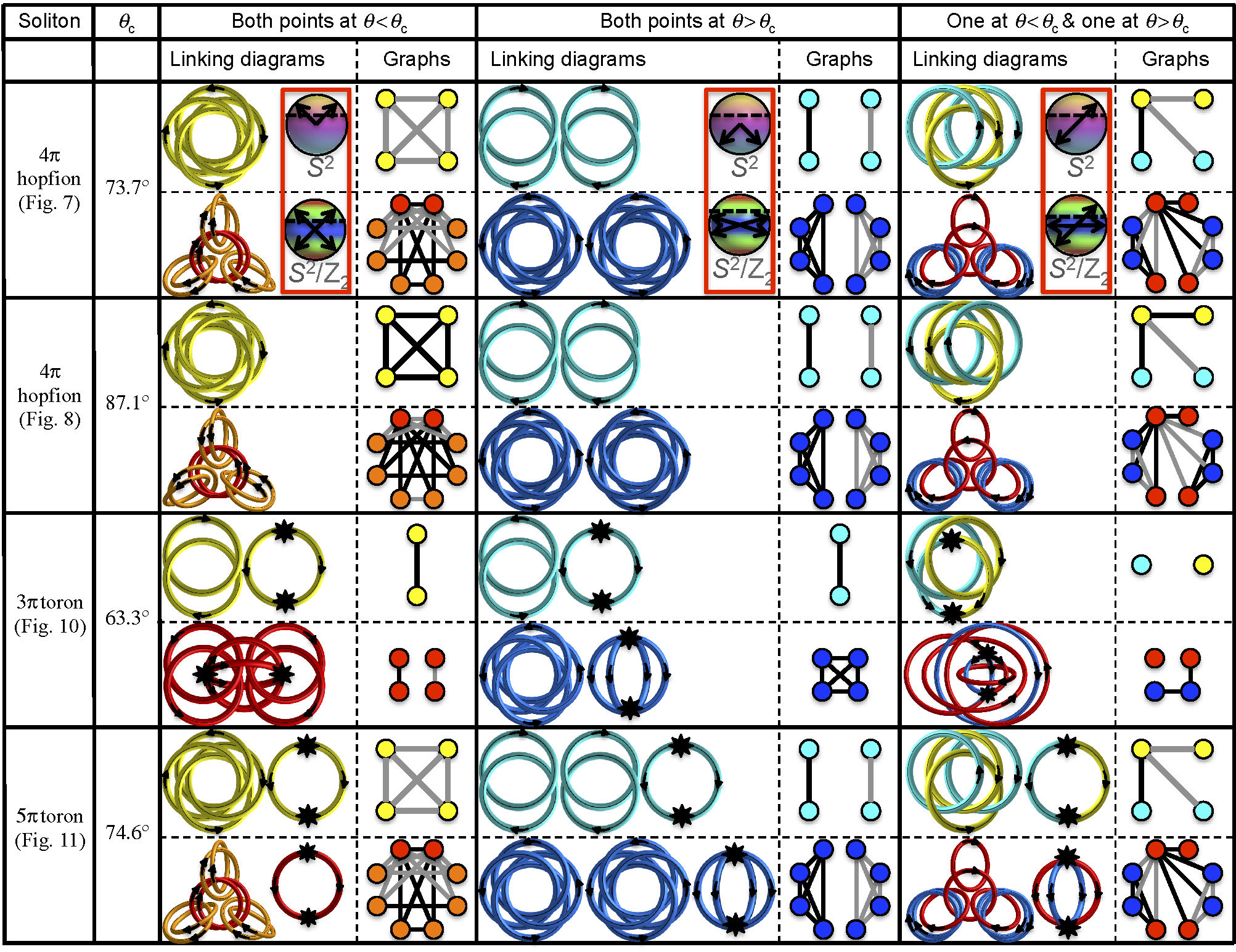}
\end{center}
Table 2. Linking diagrams and graphs of complex 3D solitons. The table presents the analysis of linking of preimages of two separate points on $S^2$ and $S^2/Z_2$ for composite 3D topological solitonic field configurations on the basis of both nonpolar and vectorized $\bf n(r)$ of the studied structures. The insets in the red boxes at the top of the columns ``linking diagrams'' depict the order parameter spaces of vectorized (top) and nonpolar (bottom) $\bf n(r)$, with arrows or double arrows indicating the points for which the preimage linking is analyzed. The dashed lines on the $S^2$ and $S^2/Z_2$ schematics separate the fragments of the $S^2$ and $S^2/Z_2$ with $\theta<\theta_c$ (top parts) and $\theta>\theta_c$ (bottom parts). The locations of the points corresponding to preimages, shown using single and double arrows on $S^2$ and $S^2/Z_2$ are the same for all solitons within the same column. In the graphs, the individual links are indicated by black or gray lines connecting the corresponding colored filled circles that represent closed-loop preimages (the black lines indicate positive signs of Linking of preimages as determined by circulations while the gray lines correspond to the negative ones). The colors of the filled circles are indicative of the points on $S^2$ (for schematics shown above the horizontal dashed lines of the table) or $S^2/Z_2$ (for schematics shown below the horizontal dashed lines of the table); for $\bf n(r)$ at $\theta<\theta_c$, two out of eight filled circles of the graphs are shown as red and the rest as orange to distinguish them on the basis of the number of times the corresponding preimages are linked. The mutually linked preimage rings in the simplified topology presentations are also shown in colors corresponding to their locations on $S^2$ or $S^2/Z_2$ and have arrows denoting circulation consistent with the far-field preimage. The point defects of torons within the topological skeletons are shown using black stars. Both the topological skeleton and graph representations of the preimage structures are constructed for the same solitonic field configurations and are provided next to each other for the case of vectorized $\bf n(r)$.
\begin{multicols}{2}

A detailed analysis of the linking diagrams (Table 2) shows that Pontryagin-Thom construction does not fully reveal the topology of complex 3D solitons, which requires directly analyzing preimages of all points on the ground-state manifold and their interlinking, not just a subset of them. Indeed, the linking numbers for the $\bf n$ and $-\bf n$ preimages forming two separate tori at $\theta>\theta_c$ (marked on the Pontryagin-Thom surfaces in Figs. 7f and 8f for $\theta=\pi/2$) change with varying $\theta$ (Table 2). The supplementary videos S1 and S2 show that this change in the linking of preimages is directly related to the transformation of the two tori corresponding to constant $\theta$-values, which occurs at $\theta_c$. This transformation is manifested by a transition from two separate concentric tori (similar to the ones shown in Figs. 7f-h and 8f-h) to two inter-nested tori like the ones depicted in Figs. 7e and 8e. The alignment and merging of preimages residing on two different tori that leads to the change of linking of constituent two rings comprising the individual preimages is interesting and calls for exploration of similar field configurations in other branches of physics. Finally, similar to the case of elementary hopfions (Figs. 4 and 5), the comparison of 3D isosurface plots of normalized handedness and free energy density (Figs. 6h,i,q,r, 7i,j and 8i,j) shows that the stability of such 3D solitons is greatly enhanced by the medium's chiral nature and tendency to form twisted structures consistent with the intrinsic pitch $p$ of the used chiral LC.

\subsection{Torons with closed-loop preimages}

In addition to the elementary torons with $\pi$-twist of $\bf n(r)$ from their central axis to the $\bf n_0$-periphery in all radial directions (Fig. 3) [13], torons with larger amounts of such twist also exist (Fig. 9a,b, 10a,b and 11a,b) [36]. For example, the torons shown in Figs. 9 and 10 have 3$\pi$- and the ones in Fig. 11 have 5$\pi$-twist of $\bf n(r)$ in all radial directions from the toron's central axis (parallel to $\bf n_0$) to the far-field periphery (Table 2), respectively. The preimages of the distinct points on the $S^2/Z_2$ or $S^2$ are either closed loops or bands terminating on the two hyperbolic point defects (Figs. 9c-e, 10c-h and 11c-h). Some of the high-twist torons exhibit re-linking of preimages (Figs. 10c-h and 11c-h and supplementary video S3 and S4), similar to that observed for the solitons shown in Figs. 7 and 8. The critical polar angles of re-linking are $\theta_c\approx63^\circ$ and $\theta_c\approx74^\circ$ for the structures shown in Figs. 10 and 11, respectively (Table 2). For the toron structure shown in Fig. 9, on the other hand, $\theta_c\approx90^\circ$ and, therefore, they effectively can be thought of as separate elementary hopfion (Figs. 4 and 5) and an elementary toron (Fig. 3) arranged coaxially in such a way that their vertical axes coincide and are parallel $\bf n_0$. The analysis of linking of closed-loop preimages for single and distinct points on $S^2/Z_2$ and $S^2$ reveals that such complex toron structures could be thought of as hybrids of elementary torons (Fig. 3) and different nonsingular solitons that we discussed above. Interestingly, the multicomponent preimages are comprised of closed loops and half-loop bands that terminate on the point singularities. The components of preimages inter-transform between one another, revealing a large diversity of torons (Table 2). These findings show an unexpected large diversity of torons and that the torons should be classified not only based on the types of the constituent self-compensating singular defects [13], but also by the types and linking of preimage components and different preimages of the nonsingular solitonic part of the torons (Table 2).   

\subsection{Twistions as composite solitonic structures}

Our method of analyzing preimages can be also applied to twistions, localized structures that embed twisted regions into a uniform background of the far-field but lack axial symmetry and (unlike torons) contain more than two self-compensating point defects [37] (Fig. 12). Although we provide here an example of a twistion with the amount of twist from its interior to periphery by $\sim$$\pi$, the analogs of twistions with larger amounts of twist in the director field can exist too and will be a subject of our future studies. The configuration of such a twistion with a stretched closed loop of $\pi$-twist of $\bf n(r)$ and four self-compensating hyperbolic point defects is shown with the help of in-plane and vertical cross-sections in Fig. 12a-c. The preimages of single points on $S^2/Z_2$ ($S^2$ for the vectorized director) are bands spanning between the four point singularities (Fig. 12d,e). This example of the twistion shows that the localized toron- and hopfion-like field configurations in confined chiral nematic LCs are not restricted to hosting none (as in hopfions) or only pairs (as in the torons) of self-compensating singular defects, but that such self-compensation can occur in a number of other more complex ways, e.g. through the co-existence of four self-compensating hyperbolic point defects shown in addition to various solitonic components with band-like or closed-loop preimages (Fig. 12). In addition to the number and types of singular defects, another source of diversity of solitonic structures can emerge from the large variety of nonsingular twisted regions and preimages that they can exhibit, which we will explore elsewhere. 

\end{multicols}
\begin{figure}[H]
\begin{center}
\includegraphics[width=0.6\textwidth]{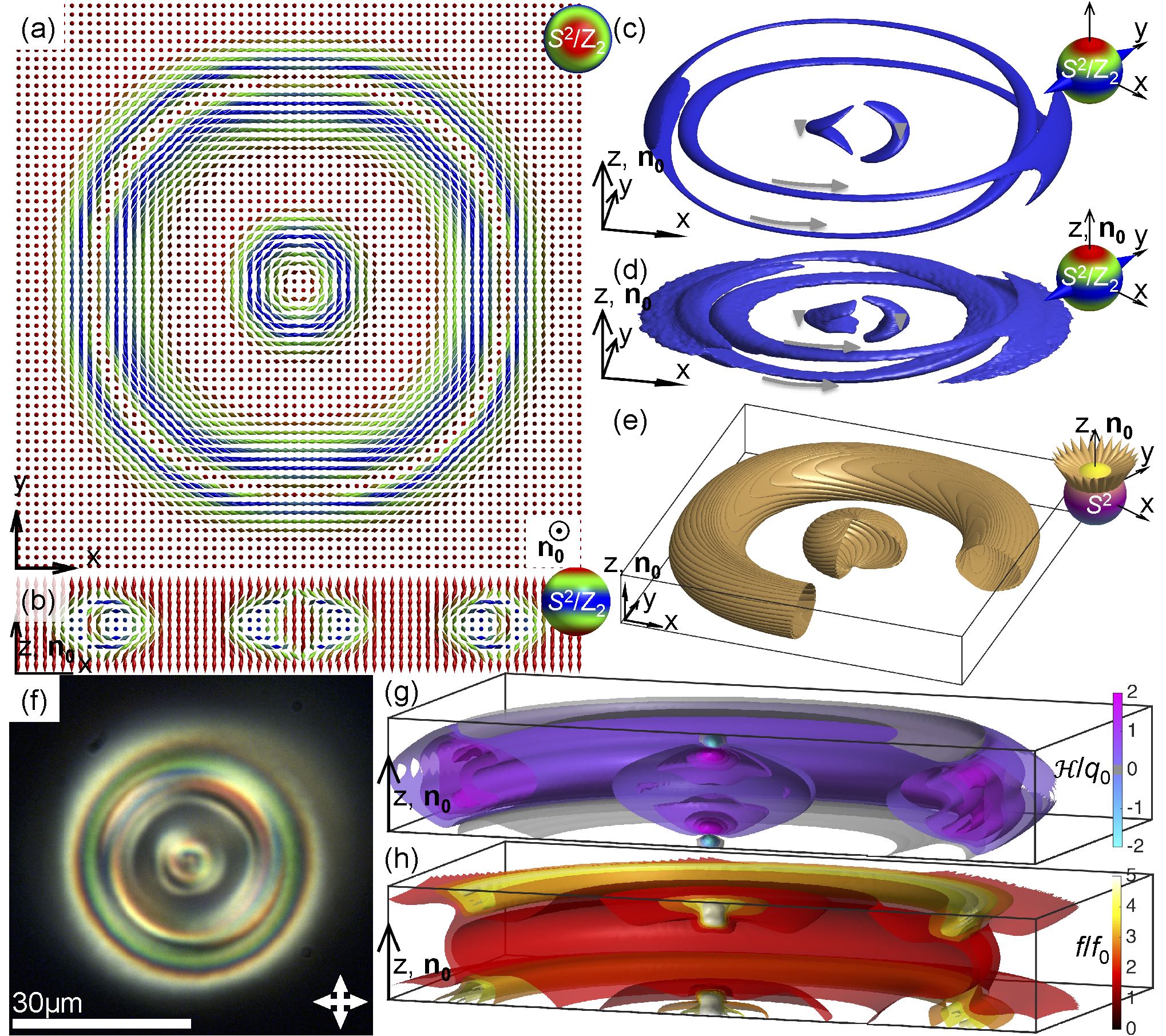}
\end{center}
\caption{3D toron-hopfion hybrid solitons formed by coaxial arrangement of an elementary toron and elementary hopfion of $Q=1$. (a,b) Computer simulated (a) in-plane orthogonal to $\bf n_0$ and (b) vertical containing $\bf n_0$ cross-sections of the axially symmetric $\bf n(r)$-structure of a soliton depicted using colored double cones; the color scheme that establishes the correspondence between director orientations and the points on $S^2/Z_2$ (top-right insets). (c) Computer-simulated and (d) experimental preimages of the soliton for the diametrically opposite points on $S^2/Z_2$ marked by double cones in the right-side insets. Gray arrows indicate the consistently determined circulations of the preimages. By analyzing such preimages of all points on $S^2/Z_2$, we find no inter-linking between the preimages of the hopfion and toron, as well as the hopfion's Hopf index $Q=1$. The 3D preimages were reconstructed based on 3PEF-PM images of these structures after the unpolymerized 5CB was replaced by immersion oil. Computer simulations were preformed for elastic constants of 5CB while also assuming $K_{24}=0$. (e) For a constant polar angle value, the closed-loop preimages of individual points on $S^2$ tile into a torus and a sphere surfaces sharing the same vertical axis parallel to $\bf n_0$, with the sphere having two small holes at the poles corresponding to the location of the hyperbolic point defects. (f) A polarizing optical micrograph of such a 3D soliton, with the white double arrows showing crossed polarizers. The polarizing optical micrograph in (f) was obtained for a structure in a 5CB-based partially polymerizable nematic LC mixture with a chiral additive CB-15 in a cell with thickness $d=10\mu$m and $d/p\sim1$. (g,h) Perspective views of the computer-simulated isosurfaces of normalized (g) handedness and (h) free energy density for this axially symmetric soliton, with the corresponding color schemes shown in the right-side insets. }
\end{figure}
\begin{multicols}{2}

\end{multicols}
\begin{figure}[H]
\begin{center}
\includegraphics[width=0.6\textwidth]{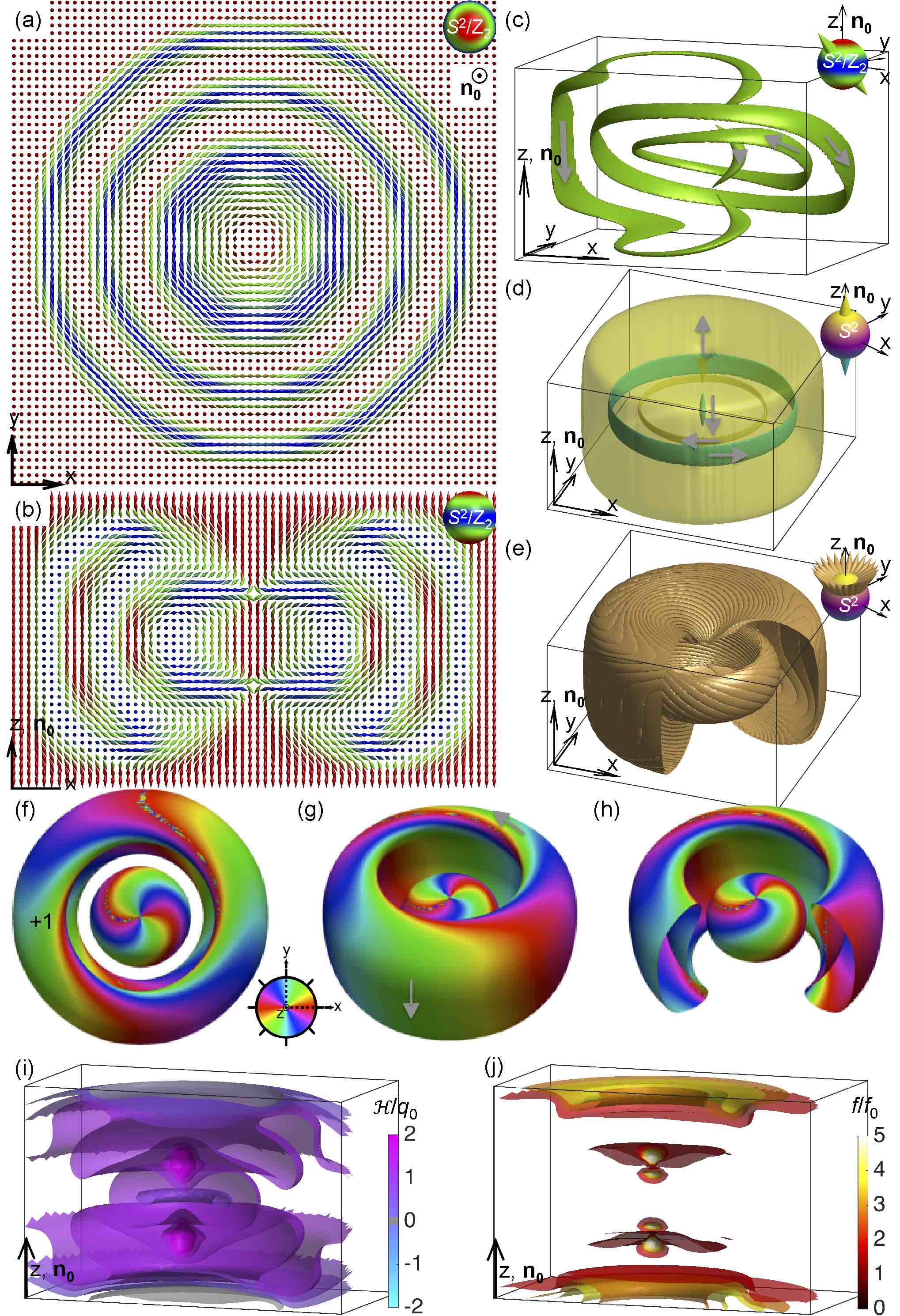}
\end{center}
\caption{Toron with complex linking and 3$\pi$ twist from its central axis to the far-field periphery. (a,b) Computer simulated (a) in-plane and (b) vertical cross-sections of the axially symmetric $\bf n(r)$-structure of the toron depicted using double cones and the color scheme that establishes correspondence between director orientations and the points on $S^2/Z_2$ (top-right insets). The two small regions of discontinuity in orientation of the double cones are the hyperbolic point defects. (c) A representative computer-simulated preimage of the hopfion for the diametrically opposite points on $S^2/Z_2$ marked by double cones in the top-right inset. The preimage is comprised of two closed loops and two half-loop bands terminating on the point defects. (d,e) Computer-simulated preimages of the toron in a vectorized $\bf n(r)$ (d) for two diametrically opposite points on $S^2$ corresponding to its north and south poles (top-right inset) and (e) for a set of points characterized by a constant polar angle $\theta$ and forming a circle parallel to the spheres equator (top-right inset). Note that the closed-loop preimages in (e) reside on a torus surface while half-loop bands form another surface spanning between the point defects. A large variety of preimages of this solitonic configuration are shown in the supplementary video S3.  (f-h) Three representative views on the isosurfaces of $\theta=\pi/2$ ($n_z=0$) colored by azimuthal orientations of $\bf n(r)$ according to the scheme shown in the right-side inset of (f). The ``-1'' on top of the torus shown in (f) indicates the linking number that characterizes the inter-linking of colored closed-loop bands and preimages of $\bf n(r)$ corresponding to points on the equator of $S^2/Z_2$. Gray arrows in (c,d) and on the green bands of (g) indicate the consistently determined circulations of the preimages. (i,j) Perspective views of the 3D computer-simulated isosurfaces of normalized (i) handedness and (j) free energy density for this axially symmetric toron, with the corresponding color schemes shown in the right-side insets. Computer simulations were preformed for elastic constants of ZLI-2806 and $d/p=2$.}
\end{figure}
\begin{multicols}{2}

\end{multicols}
\begin{figure}[H]
\begin{center}
\includegraphics[width=0.6\textwidth]{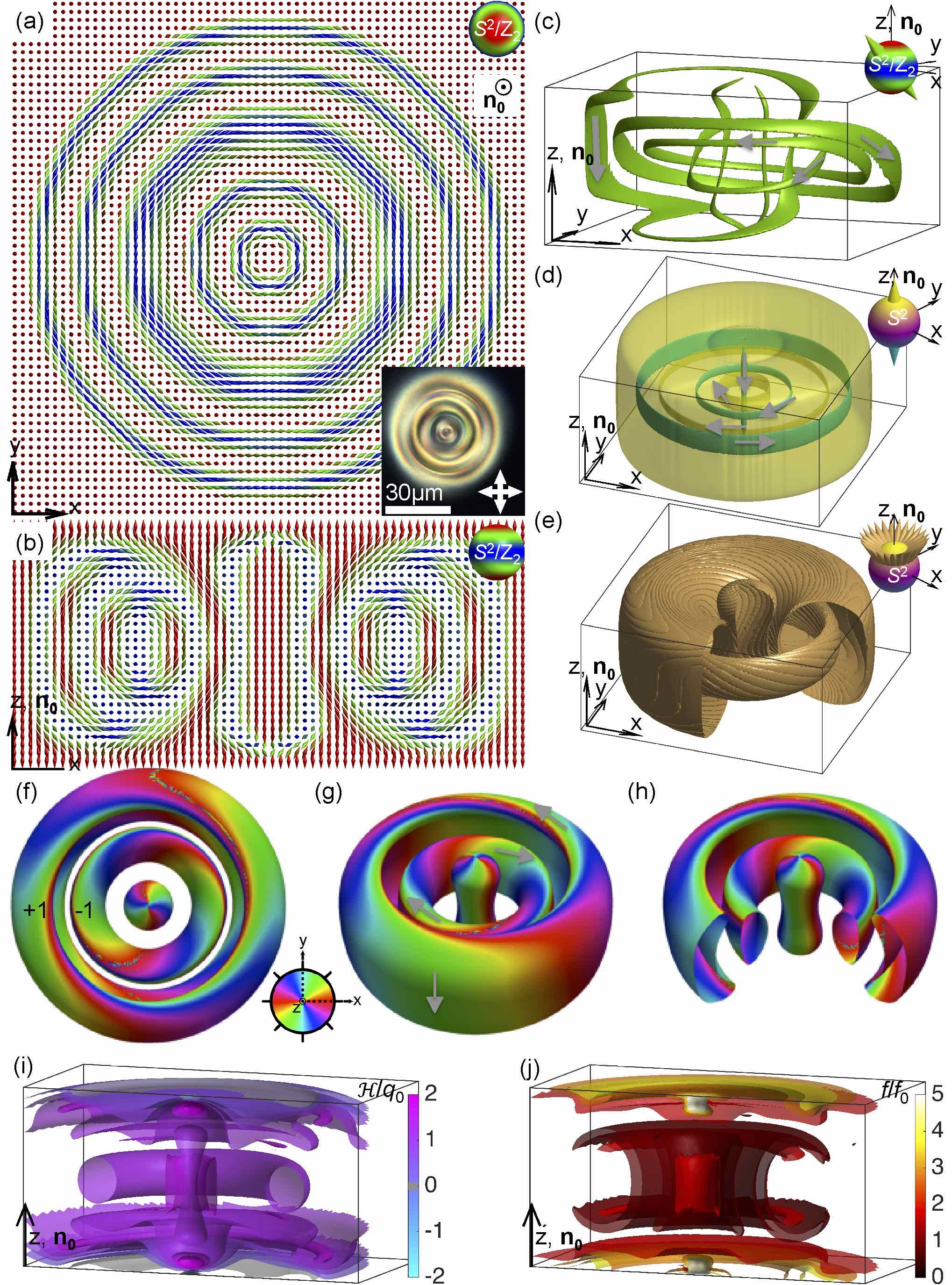}
\end{center}
\caption{Toron with complex linking and 5$\pi$ twist from its central axis to the far-field periphery. (a,b) Computer simulated (a) in-plane and (b) vertical cross-sections of the axially symmetric $\bf n(r)$-structure of the toron depicted using double cones and the color scheme that establishes correspondence between director orientations and the points on $S^2/Z_2$ (top-right insets). The two small regions of discontinuity in orientation of the double cones are the hyperbolic point defects. The inset in (a) shows a polarizing optical micrograph of such a structure. The optical micrograph was obtained for a 5CB-based partially polymerizable LC in a cell of $d=10\mu$m.  (c) A representative computer-simulated preimage of the hopfion for the diametrically opposite points on $S^2/Z_2$ marked by double cones in the top-right inset. The preimage is comprised of four inter-linked closed loops and two half-loop bands terminating on the point defects. (d,e) Computer-simulated preimages of the toron in a vectorized $\bf n(r)$ (d) for two diametrically opposite points on $S^2$ corresponding to its north and south poles (top-right inset) and (e) for a set of points characterized by a constant polar angle $\theta$ and forming a circle parallel to the spheres equator (top-right inset). Note that the closed-loop preimages in (e) reside on two nested tori surfaces while half-loop bands form another surface spanning between the point defects. A large variety of preimages of this solitonic configuration are shown in the supplementary video S4. (f-h) Three representative views on the isosurfaces of $\theta=\pi/2$ ($n_z=0$) colored by azimuthal orientations of $\bf n(r)$ according to the scheme shown in the right-side inset of (f). The numbers on top of the tori shown in (f) indicate the linking numbers that characterize the inter-linking of colored closed-loop bands and preimages of $\bf n(r)$ corresponding to points on the equator of $S^2/Z_2$. Gray arrows in (c,d) and on the green bands of (g) indicate the consistently determined circulations of the preimages. (i,j) Perspective views of the 3D computer-simulated isosurfaces of normalized (i) handedness and (j) free energy density for this axially symmetric 3D soliton, with the corresponding color schemes shown in the right-side insets. Computer simulations were preformed for elastic constants of ZLI-2806 and $d/p=2$.}
\end{figure}
\begin{multicols}{2}

\end{multicols}
\begin{figure}[H]
\begin{center}
\includegraphics[width=0.6\textwidth]{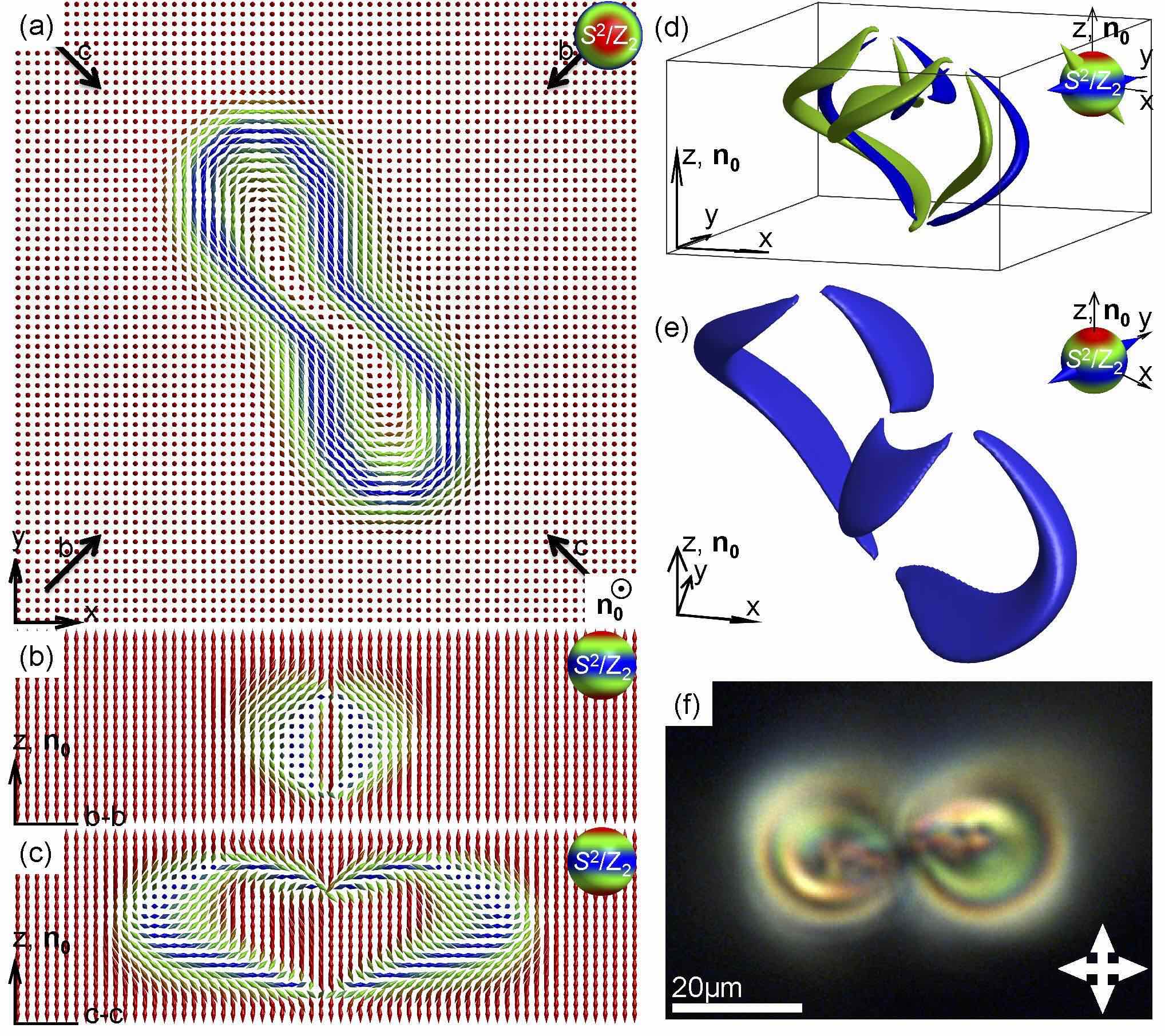}
\end{center}
\caption{A twistion structure in a chiral nematic LC. (a-c) Computer simulated (a) in-plane and (b,c) vertical cross-sections of the 3D $\bf n(r)$-structure of the twistion depicted using double cones and the color scheme that establishes correspondence between director orientations and the points on $S^2/Z_2$ (top-right insets). The locations of vertical cross-sections (b) and (c) are depicted in (a) using arrows. (d) Computer-simulated preimages of the twistion for two sets of the diametrically opposite points on $S^2/Z_2$ marked by double cones in the top-right inset. (e) Computer-simulated preimages of the twistion for points on the ``equator'' of $S^2/Z_2$ (top-right inset). Computer simulations were preformed for elastic constants of 5CB and $d/p=0.85$. (f) A polarizing optical micrograph of the twistion, with the white double arrows showing crossed polarizers. The polarizing micrograph was obtained for a structure in a 5CB-based partially polymerizable nematic mixture doped with a chiral additive CB-15 in a cell with $d/p\sim1$.}
\end{figure}
\begin{multicols}{2}

\section{Discussion}

Although the 3D solitons with nonzero Hopf invariants are theoretically predicted to exist in many branches of science, ranging from particle physics to cosmology, their experimental identification and detailed study is often prohibitively difficult. Even in the case of solid-state chiral ferromagnets, in which the two-dimensional counterparts of the 3D solitons are recently extensively studied [25-29], experimental imaging techniques are lacking the ability to resolve details of field configurations within the nanometer-sized localized structures with high resolution in 3D. So, in fact, the 3D topological solitons may (under certain circumstances, considering the similar description by Eq. (2) within the simples models [24,25]) exist in the solid-state chiral ferromagnet systems but the lack of appropriate imaging and analysis techniques prohibits their experimental identification and classification. Chiral nematic LCs provide an experimental advantage of hosting micrometer-sized 3D solitonic structures, so that their structure is accessible to the direct 3D nonlinear optical imaging [35]. The experimental 3D configurations of $\bf n(r)$ within the studied solitons closely agree with numerical modeling, allowing us to robustly identify and classify the nonsingular $\pi_3(S^2/Z_2)=\mathbb{Z}$ topological defects with different Hopf indices. This will provide important insights needed for the realization of topological solitons in other physical systems. We also envisage that chiral LCs will serve as a test bed for theories of 3D topological solitons.

The possibility of realizing 3D localized field configurations embedded in a uniform far-field background as static solitons is a subject of active studies in different branches of theoretical physics and applied mathematics [3]. The Hobart-Derrick theorem states that the static 3D solitons cannot have finite energy for the free energy functional resembling the first term in Eq. (2) [7,8]. Indeed, our numerical modeling confirms that all computer-simulated 3D solitons discussed above become unstable after removing the chiral terms of free energy in Eqs. (1) and (2) for the nonchiral nematic LC with $q_0=0$ while using all other parameters within the experimentally accessible ranges, consistent with the corresponding experiments at otherwise identical conditions. However, the chiral LC medium's tendency to twist $\bf n(r)$ in the frustrated confined geometry of unwound homeotropic cells helps to embed energetically favorable twisted regions of solitons into the uniform unwound background of $\bf n_0$. When the LC cell thickness $d$ is comparable to the intrinsic helicoidal pitch of the chiral nematic medium, with $d/p=0.5-2$, a large number of spatially localized structures with twisted $\bf n(r)$ can become embedded in the uniform far-field background (Figs. 2-12) to locally relieve the frustration imposed by the incompatibility of homeotropic boundary conditions and the helicoidal structure of the chiral LC's ground-state. Interestingly, all of the studied 3D solitons emerge as local minima of the free energy functionals given by both Eqs. (1) and (2), albeit at somewhat different $d/p$ ratios and other parameters. The analysis of 3D isosurfaces of the twist handedness $\mathcal{H}$ reveals how LC's chirality mediates appearance of the twisted solitons with finite dimensions comparable to $p$ by showing that $\mathcal{H}$ within the localized structures is mostly the same as that of the ground-state chiral nematic LC in which they are hosted. However, an interesting, unexpected observation is that the localized structures also posses small energetically costly regions with $\mathcal{H}$ opposite to that of the ground-state LC medium. This finding may imply that the twisted solitons require reversal of $\mathcal{H}$ to match their internal field configurations with the uniform far-field $\bf n_0$, albeit this aspect will require separate detailed studies. 

A comparison of Eqs. (1) and (2) helps to identify elastic constant anisotropy as (unique to LC systems) an additional mechanism for stabilizing the 3D knotted solitons. Indeed, although we find that the studied solitons and torons are solutions of both Eqs. (1) and (2), the parameter space of stability is different. For example, the studied structures of elementary hopfions (Figs. 4 and 5) tend to be stable at smaller $d/p$ values roughly within $0.8-1$ for the case of one-elastic-constant approximation but are found to emerge as solutions of Eq. (1) for the elastic constants of 5CB roughly within $0.95-1.2$ when we set $K_{24}=0$. In addition to the chirality and elasticity, the stability of 3D solitonic structures can likely be further controlled by varying the strength of vertical surface boundary conditions for the director and by applying external fields that would dielectrically or diamagnetically couple to $\bf n(r)$. For example, in the case of LCs with positive dielectric anisotropy, the applied low-voltage fields ($1-5$ V) and the electric field coupling term of the free energy given by Eq. (3) tends to extend the range of stability of elementary hopfions to larger $d/p$ values. Since the focus of our study is on topology of the 3D solitonic structures, the exploration of detailed structural diagrams of hopfion and toron stability is beyond the scope of this work, but will be a subject of our future studies.

An interesting observation is that fully nonsingular solitons in $\bf n(r)$ emerge when the twist of director from the central axis of these axially symmetric structures to the periphery is an even integer of $\pi$ (2$\pi$ and 4$\pi$ in the provided examples in Figs. 4-8) while several different torons with point singularities have an odd number of $\pi$ of such twist ($\pi$, 3$\pi$ and 5$\pi$ in the provided examples in Figs. 3,9-11). This observation provides insights into one of the sources of diversity of studied solitonic structures in confined chiral LCs, which stems from the amount of the director twist in radial directions of the axially symmetric structures of solitons, as well as insights into how such structures can be generated on demand experimentally. Another interesting observation is that the far-field distortions of $\bf n(r)$ corresponding to most of the studied solitonic configurations are symmetric with respect to the sample midplane, a plane crossing centers of the localized structures orthogonally to $\bf n_0$, except for the topologically trivial soliton with $Q=0$ shown in Fig. 2 and the twistion structure presented in Fig. 12. These structural features can be analyzed based on vertical cross-sections of the field configurations in the planes containing $\bf n_0$ (Figs. 1-12) and are related to the multitude of different ways of embedding localized twisted regions within the uniform far-field background $\bf n_0$. Consistently with the peculiarities of the up-down symmetry of the solitonic configurations in the homeotropic cells, we find that the solitons exhibit richness of elastic interactions and self-assembly. We previously explored such interactions and self-assembly for the elementary torons [36], but will also extend such studies to the other solitonic structures, which will be reported elsewhere.

We have demonstrated that our method of preimage analysis provides a comprehensive way of exploring the topology of 3D solitons, providing insights into the nature of both nonsingular topologically nontrivial structures and the singular defects such as the point singularities found in the studied torons. This approach is further expanding the capabilities of the method of Pontryagin-Thom construction that we also used for this purpose [14]. A detailed analysis of preimages allows us to uncover a number of rather unexpected features of 3D solitons in LCs. For example, the 3D soliton with $Q=0$ shown in Fig. 2 has closed-loop preimages for the majority of $S^2$-points (for the vectorized director field), except for the vicinity of the south pole of the $S^2$-sphere. This information could have been missed had we not analyzed preimages of all $S^2$-points but only some of them, as in the case of the Pontryagin-Thom construction. A comparison of topological linking of preimages of the solitons shown in Figs. 7 and 8 based on the summary presented in Table 2 additionally emphasizes the need to analyze both vectorized and nonpolar $\bf n(r)$. Indeed, the different topological nature of these two solitons could not be revealed based on the Pontryagin-Thom constructions, which are rather similar for the two structures (compare Fig. 7f-h and Fig. 8f-h). Moreover, we find exactly the same linking of the preimages of all pairs of points on $S^2$ and $S^2/Z_2$ (Table 2 and Figs. 7 and 8) of these two different solitonic structures and the difference between them can be seen only when the preimage circulations are consistently defined. For both nonpolar and vectorized $\bf n(r)$, we can see the difference between topologies of these two solitons on the basis of circulation directions (Table 2 and Figs. 7 and 8), with the differences in preimage linking apparent when one $S^2$-point is at $\theta<\theta_c$ and one at $\theta>\theta_c$ or both are at $\theta<\theta_c$, but not when both of the analyzed points are at $\theta>\theta_c$. In principle, 3D solitons with different structures could have the same topology of preimage linking, being homeomorphic to each other, but this is not the case for the solitons shown in Figs. 7 and 8, which cannot be smoothly morphed one to another. The detailed analysis of preimages allows us to identify and demonstrate such subtle differences between the 3D solitons. In a similar way, the important differences between the four different types of torons that we present in this study could be missed without a detailed analysis of preimages of all points on $S^2/Z_2$ for nonpolar and on $S^2$ for vectorized $\bf n(r)$ (Figs. 3 and 9-11 and Table 2). On the other hand, the analysis of only the closed-loop preimages but not the ones terminating on point defects would fail to reveal differences between certain types of torons and solitons without point defects (compare Fig. 7 and Fig. 11, as well as the corresponding summaries presented in Table 2). 

The 3D solitons that we discuss in this work constitute a non-exhaustive, illustrative set of examples of topologically nontrivial field configurations that can be stabilized in chiral liquid crystals and ferromagnets, but a much larger variety of $\pi_3(\mathbb{R}P^2) \equiv \pi_3(S^2/Z_2)=\mathbb{Z}$ and $\pi_3(S^2)=\mathbb{Z}$ topological defects can exist in these condensed matter systems and will be the subject of our future studies. Among many interesting questions that remain to be answered, one concerns finding the different ways in which various stable and metastable states with the same topology (and $Q$) can be realized in the studied system. For example, the uniform unwound state and the 3D solitons shown in Figs. 2 and 3 are all characterized by $Q=0$, but it remains to be found whether other solitons with $Q=0$ can be realized experimentally and as local free energy minima in modeling. Importantly, our approach of imaging preimages is ideally suited to reveal the diversity and complexity of the 3D solitons in LCs. Indeed, we note that despite the fact that polarizing optical micrographs (Figs. 1f,g, 6c,l, 9f, 11a, and 12f) of different solitons differ from each other, they do not allow for determining the type of preimages and their interlinking, which is only possible to do on the basis of the 3D nonlinear optical imaging and with the help of the method of preimages that we have introduced. 

Although 3D topological solitons have been studied as part of dynamic and transient phenomena in many different physical systems (for example, see [14,16,22,48]), chiral nematic LCs are perhaps the only system in which these solitons are realized as long-term stable configurations accessible to detailed experimental studies of their 3D structure and topology, which makes them ideally suited to serve as model systems for the study of $\pi_3(S^2/Z_2)=\mathbb{Z}$ and $\pi_3(S^2)=\mathbb{Z}$ topological defects. For example, the initial interest in 3D topological solitons emerged in the fields of particle physics and cosmology [1, 3, 6-10, 48], where they continue to play important roles [3]. Although the Skyrme model (and related models such as the Skyrme-Faddeev model [9]) was initially proposed as a model describing strong interactions of hadrons [6], Witten and colleagues later demonstrated that similar ideas could be derived on the basis of quantum chromodynamics (QCD) [49]. In the low-energy pion dynamics model, certain elementary particles (including protons) can be thought of as $\pi_3$ textures/solitons [48,49,50]. In addition, the 3D topological solitons are also predicted to occur in cosmology [48,50] and in many other physical systems [3], in which their detailed study is typically inaccessible to the direct experimentation. We thus expect that the use of chiral nematic LCs as a model system to study 3D topological solitons may impinge on their understanding in contexts of physics phenomena in other branches of physics. Furthermore, since the solitonic structures with different Hopf indices are topologically distinct from each other, transformations between them are discontinuous and involve energetic barriers. Thus, different types of solitonic structures can be obtained as long-lived states corresponding to locally different optical and other properties. Since the different solitons incorporate different patterns of the effective refractive index distribution, they could serve as means of realizing reconfigurable phase gratings [44], pixels for bi-stable and multi-stable displays, etc. If the 3D topological solitons discussed here can be also discovered in solid ferromagnets [24-29], they can potentially revolutionize the field of Skyrmionics currently enabled by their two-dimensional counterparts, the so called ``baby skyrmions'' [28,29].

To conclude, we have introduced an approach for experimental and numerical analysis of 3D topological solitons with nonzero Hopf invariants. Within this approach, inspired by the mathematical Hopf maps, point-by-point, we experimentally scan the order parameter space (the $S^2$ sphere or $S^2/Z_2$) and find regions within the sample that have orientations of the director/vector corresponding to the point of $S^2$ or $S^2/Z_2$. The same procedure was implemented numerically based on field configurations arising from the minimization of free energy, allowing for the unambiguous characterization of the topology of 3D solitons. We have applied this analysis of experimental and numerical preimages and Hopf maps as a means of uncovering an unexpectedly large diversity of 3D spatially localized solitonic structures in confined chiral nematic LCs. We have revealed a host of torons, hopfions, and other solitons with complex linking of closed-loop preimages and both with and without singular point defects. Self-assembly of such 3D solitons with different topological characteristics may result in emergence of new condensed matter phases with rich phase diagrams and unusual physical behavior. A comparison of nematic and ferromagnetic hopfions and torons that we studied recently [22] will allow for probing the role of field polarity in the topology of knotted solitons. Finally, the experimental and theoretical frameworks that we have introduced will help establishing chiral nematic LCs as a test-bed for the study of 3D topological solitons, which are abundant in theories in practically all branches of physics.

\section{Acknowledgments}
~~~ We are grateful to D. Broer for providing the RM-82 and RM-257 reactive mesogens. We acknowledge discussions with S. de Alwis, T. DeGrand, N. Clark, R. Kamien, T. Lubensky, H. Osman, and M. Tasinkevych, as well as the support of the US National Science Foundation Grant DMR-1410735.

\end{multicols}

\end{document}